\documentclass[10pt, twocolumn]{IEEEtran}

\usepackage{times,amsmath}
\usepackage{amsfonts}
\usepackage{amssymb,bm}
\usepackage{amsthm}
\usepackage{algorithm}
\usepackage{algpseudocode}
\usepackage{cite}
\usepackage{graphicx}
\usepackage{epstopdf}
\usepackage[caption=false,font=footnotesize]{subfig}
\usepackage{fixltx2e}
\usepackage{balance}
\usepackage{cases}

\usepackage{color}

\begin{document}
%
\title{Robust Transmission in Downlink Multiuser MISO Systems: A Rate-Splitting Approach}
\author{Hamdi~Joudeh,~\IEEEmembership{Student Member,~IEEE,} and
        Bruno~Clerckx,~\IEEEmembership{Member,~IEEE}
\thanks{Copyright (c) 2015 IEEE. Personal use of this material is permitted. However, permission to use this material for any other purposes must be obtained from the IEEE by sending a request to pubs-permissions@ieee.org.}
\thanks{This work is partially supported by the UK Engineering and Physical Sciences Research Council (EPSRC) under grant EP/N015312/1.
This paper was presented in part at the 41st IEEE International Conference on Acoustics,
Speech and Signal Processing (ICASSP), Shanghai, China, March 2016 \cite{Joudeh2016}.}
\thanks{The authors are with the Communications and Signal Processing Group,
Department of Electrical and Electronic Engineering, Imperial College, London SW7 2AZ, U.K. (email: hamdi.joudeh10@imperial.ac.uk; b.clerckx@imperial.ac.uk). B.~Clerckx is also with the School of Electrical Engineering, Korea University, Seoul 136-713, Korea.}
}
\maketitle

\begin{abstract}
We consider a downlink multiuser MISO system with bounded errors in the Channel State Information at the Transmitter (CSIT).
We first look at the robust design problem of achieving max-min fairness amongst users (in the worst-case sense).
Contrary to the conventional approach adopted in literature, we propose a rather unorthodox design based on a Rate-Splitting (RS) strategy.
Each user's message is split into two parts, a common part and a private part.
All common parts are packed into one \emph{super} common message encoded using a public codebook, while private parts are independently encoded.
The resulting symbol streams are linearly precoded and simultaneously transmitted, and each receiver retrieves its intended message by decoding both the common stream and its corresponding private stream.
For CSIT uncertainty regions that scale with SNR (e.g. by scaling the number of feedback bits),
we prove that a RS-based design achieves higher max-min (symmetric) Degrees of Freedom (DoF) compared to conventional designs (NoRS).
For the special case of non-scaling CSIT (e.g. fixed number of feedback bits), and contrary to NoRS, RS can achieve a non-saturating max-min rate.
We propose a robust algorithm based on the cutting-set method coupled with the Weighted Minimum Mean Square Error (WMMSE) approach, and we demonstrate its performance gains over state-of-the art designs.
Finally, we extend the RS strategy to address the Quality of Service (QoS) constrained power minimization problem, and we demonstrate significant gains over NoRS-based designs.
\end{abstract}


\IEEEpeerreviewmaketitle
\section{Introduction}
\label{Section_Introduction}
\newcounter{Proposition_Counter} 
\newcounter{Theorem_Counter} 
\newcounter{Lemma_Counter} 
\newcounter{Remark_Counter} 
\newcounter{Assumption_Counter}
\newcounter{Corollary_Counter}
Consider a Multiuser (MU) Multiple-Input-Single-Output (MISO) system operating in Downlink (DL), where a Base Station (BS) equipped with an antenna array sends independent messages to a number of single-antenna receivers.
In such systems, it is necessary to perform preprocessing at the BS to mitigate the channel interference and realize the high spectral efficiencies promised by employing multiple antennas/users \cite{Clerckx2013}.
Among the different preprocessing techniques, linear precoding (or beamforming) strategies have emerged as the most popular choices due to the tractability of the corresponding design problems.
Two typical design problems are:
\begin{itemize}
  \item Maximizing the minimum Quality of Service (QoS) among users subject to a total transmit power constraint.
  \item Transmit power minimization subject to QoS constraints.
\end{itemize}
The former is known as the max-min fairness problem (or balancing problem), while the latter is known as the QoS problem. The two problems are closely related: there is an inverse relation between the two when all users request the same QoS in the second problem.
In the presence of perfect Channel State Information (CSI), it has been shown that the two problems assume Signal to Interference plus Noise Ratio (SINR) formulations, and can be solved globally and efficiently using means of conic programming \cite{Bengtsson2001,Wiesel2006}.
In practical systems, the CSI at the Transmitter (CSIT) is subject to various sources of imperfection, such as estimation errors in Time Division Duplex (TDD) systems \cite{Hassibi2003}, and quantization errors in Frequency Division Duplex (FDD) systems \cite{Love2008}.
Both cases are prone to delays and mismatches arising from using outdated CSIT.
Such imperfections are known to be detrimental to the performance of naive designs that assume perfect CSIT \cite{Jindal2006,Ding2007,Caire2010}.
This motivated a wide range of robust designs which aim to guarantee a certain performance under CSIT imperfections
\cite{Mutapcic2007,Vucic2009a,Vucic2009,Song2012,Shenouda2007,Shenouda2009,Shenouda2008,Wang2014}.
The formulation of the robust optimization problem is heavily influenced by the adopted error model.
In this work, we consider the case where CSIT errors are confined within some known bounded uncertainty sets.
This is typical in scenarios where errors emerge as a result of quantization and limited channel feedback. Knowledge of the quantization codebooks can be used to bound such errors.
This model can also be applied to control the outage probability  for unbounded estimation errors \cite{Wang2014}.

Robust optimization is carried out in the worst-case sense \cite{Kassam1985}, where precoders are designed such that a certain performance is guaranteed for all possible channels in the corresponding uncertainty regions.
The robust fairness and QoS problems are non-convex in general with infinitely many constraints, and usually very difficult to solve in their raw forms.
To overcome these difficulties, several approximations have been proposed to transform the problems into tractable forms
\cite{Shenouda2007,Mutapcic2007,Vucic2009,Vucic2009a,Shenouda2009,Song2012}.
The different approaches vary in the degree of conservatism and computational complexity.
However, all existing works consider a conventional transmission scheme, i.e. each message is encoded into an independent data stream, then all streams are spatially multiplexed through linear precoding.
Optimum max-min fair designs ultimately achieve balanced rates, requiring a simultaneous increase in users' powers.
This is known to limit the performance under CSIT errors that do not decay with increased Signal to Noise Ratio (SNR), e.g. fixed number of feedback bits \cite{Jindal2006}.
Rates saturate at high SNRs where MU interference becomes dominant, and cannot be completely dealt with due to CSIT imperfection.
While robust designs enhance the performance, the inherent interference cannot be eliminated, and the rate saturates as observed in \cite{Shenouda2007}.
Conversely, this creates a feasibility issue for the power minimization problem, since rates beyond the saturation level cannot be achieved.
We primarily focus on the max-min fairness problem through this work, before extending the developed methods to the QoS problem.

To overcome the limitations of conventional designs, we propose a novel robust strategy based on Rate-Splitting (RS).
A given message is split into two parts: a common part and a private part.
All common parts are packed into one \emph{super} common message, encoded into a codeword drawn from a common codebook shared by all users.
The private part is encoded using a private codebook, known to the corresponding user.
For a $K$-user system, the resulting $K+1$ encoded streams are linearly precoded and simultaneously transmitted.
At each receiver, the common message is decoded first by treating all private signals as noise.
This is followed by decoding the private message after removing the common message via Successive Interference Cancellation (SIC).
The original messages are delivered given that each receiver decodes the common message and its private message successfully.

The idea of RS dates back to \emph{Carleial}'s work on the Interference Channel (IC), then later appears in the famous  \emph{Han-Kobayashi} scheme,  where it was shown that decoding part of the interference (in the form of common messages) enhances the performance \cite{ElGamal2011}.
The inherent MU interference in the MISO Broadcast Channel (BC) with imperfect CSIT draws a strong resemblance to the IC, one that has been generally overlooked until recently.
It was shown in \cite{Yang2013} that RS boosts the sum Degrees of Freedom (DoF) of the MISO-BC under imperfect CSIT,
where errors decay with increased SNR at a rate of $O(\mathrm{SNR}^{-\alpha})$ for some constant
$\alpha \in [0,1]$ (see also \cite{Hao2013}).
Building upon this result, the sum-rate performance in the context of limited channel feedback using basic closed-form precoders is analysed in \cite{Hao2015}, where RS is shown to outperform conventional transmission without rate-splitting (NoRS).
Maximizing the sum-rate performance of the RS scheme through precoder optimization was considered in \cite{Joudeh2015,Joudeh2016a}.
RS was also extended to FDD massive MIMO systems in \cite{Dai2016}.
From a design perspective, sum-rate (or sum-DoF) maximization in the aforementioned works is achieved by splitting any of the messages.
The RS sum-rate question is posed as whether splitting is required or not, and in case the answer is positive, how much of the total information should be relayed by the common message, regardless of which user message(s) is split.
This does not take into account individual rates and fairness amongst users.
Generalizing the RS scheme such that the common stream is shared between users to achieve fairness was introduced in our previous works \cite{Joudeh2015a,Joudeh2016a}, where average rates were optimized under statistical CSIT uncertainty.
A key feature that differentiates the RS max-min fairness design problem from its NoRS counterpart is the sum-rate nature of each user's achievable rate arising from splitting the messages.
The non-convex coupled sum-rate expressions motivate invoking the Weighted Minimum Mean Square Error (WMMSE) approach in \cite{Christensen2008,Shi2011}.
This unveils a block-wise convex structure that can be exploited using the Alternating Optimization (AO) principle.
However, the sampling-based method used to approximate the average rates in \cite{Joudeh2015a,Joudeh2016a} cannot be applied to the worst-case formulation with infinitely many constraints considered here.
In a preliminary version of this paper \cite{Joudeh2016}, we resort to the robust WMMSE algorithm proposed in \cite{Tajer2011}.
This approach first applies a conservative approximation, which then enables the abstraction of the infinitely many constraints into a finite number of linear matrix inequalities, using the $\mathcal{S}$-lemma based result in \cite{Eldar2004} (also used in \cite{Vucic2009,Vucic2009a}).
Nevertheless, it was observed in \cite{Joudeh2016} and \cite{Tajer2011}  that this approach fails to achieve the theoretically predicted DoF, which calls for an algorithm that can achieve the anticipated performance.
Before we proceed, we emphasize that for the $K$-user system considered in this work, the $k$th user's CSIT uncertainty region is modeled as a ball of radius $\delta_{k}$.
To incorporate scenarios where CSIT errors decay with increased SNR \cite{Davoodi2014}, e.g. by scaling the number of feedback bits \cite{Jindal2006}, we allow $\delta_{k}^{2}$ to scale as $O(\mathrm{SNR}^{-\alpha_{k}})$ for some CSIT scaling (or quality) factor $\alpha_{k} \in [0,1]$.
Without loss of generality, we assume that $\alpha_{1} \leq  \alpha_{2} \leq \ldots \leq \alpha_{K}$.
Next, the key contributions of this paper are listed:
\begin{itemize}
\item We characterize the optimum performances of NoRS and RS max-min fairness designs in the interference limited regime by deriving their optimum max-min DoFs\footnote{This is also known as the symmetric DoF in litrature\cite{Yang2013}, i.e. the DoF that can be achieved by all users simultaneously.} shown to be
    $\bar{d}^{\ast} = \frac{\alpha_{1}+\alpha_{2}}{2}$ and
    $\bar{d}^{\ast}_{\mathrm{RS}} = \underset{J \in \{2,\ldots,K\}}{\min} \frac{1+\sum_{k=1}^{J-1} \alpha_{k}}{J}$ respectively.
    This gives insight into the RS  performance gains, e.g. $\bar{d}^{\ast} = 0$ under $\alpha_{1},\alpha_{2} = 0$ yielding a saturating NoRS rate, while RS achieves non-saturating rates regardless of the CSIT scaling as $\bar{d}^{\ast}_{\mathrm{RS}} \geq \frac{1}{K}$.
 \item We derive a performance upper-bound for the conservative robust WMMSE approach
 \cite{Joudeh2016,Tajer2011} from which its behavior is explained.
We show that the employed approximation introduces self-interference
 terms that undermine the worst-case achievable rates, as perceived and guaranteed by the BS, specifically at high SNRs.
Such limitations were not identified previously.
 \item We propose an algorithm based on the cutting-set method \cite{Mutapcic2009}, which solves the problem by alternating between an optimization step: where a solution is obtained w.r.t a finite subset of the uncertainty region, and a pessimization step: where the subset is updated through worst-case analysis.
     Contrary to works adopting this method \cite{Mutapcic2007,Ubaidulla2011,Ubaidulla2014,Liu2015}, we avoid conservative approximations used to convexify the optimization step, and we solve the pessimization step exactly, and hence guarantee convergence to a robust solution of the original semi-infinite problem. We further prove the KKT optimality of the solution.
  \item We show through simulations that for a given transmission scheme (RS or NoRS), the proposed cutting-set method is superior to the conservative method of \cite{Tajer2011}. On the other hand, for a given robust design method, RS provides significant performance gains over NoRS.
  \item Finally, we formulate a RS version of the QoS problem and show that the proposed algorithm resolves the feasibility issue arising in NoRS designs, and requires less transmission power to meet the same QoS constraints.
\end{itemize}
The rest of the paper is organized as follows. The system model is described in Section \ref{Section_System_Model}.
The problem is formulated in Section \ref{Section_Problem_Statement}, where the asymptotic performance is also derived.
The conservative robust WMMSE approach and its limitations are discussed in Section \ref{Section_Conservative}.
In Section \ref{Section_Cutting_Set}, we propose the cutting-set method and establish its convergence to a KKT point.
Simulations results, analysis and the QoS problem are presented in Section \ref{Section_Simulation_Results}, and Section \ref{Section_Conclusion} concludes the paper.

\emph{Notation}: Boldface uppercase, boldface lowercase and standard letters denote matrices, column vectors, and scalars, respectively.
The superscrips $(\cdot)^{T}$, $(\cdot)^{H}$ and $(\cdot)^{\dag}$ denote the transpose, conjugate-transpose (Hermitian), and pseudo-inverse operators, respectively. $\text{tr}(\cdot)$, $\|\cdot\|$ and $\mathrm{E}\{\cdot\}$ are the trace, Euclidian norm, and expectation operators, respectively. $\mathbf{A} \succeq 0$ means that the symmetric matrix $\mathbf{A}$ is positive semidefinite.
%
%
\section{System Model}
\label{Section_System_Model}
Consider a BS equipped with $N_t$ antennas serving a set of single-antenna users $\mathcal{K} \triangleq \{1,\ldots,K\}$, where $K \leq N_{t}$.
For a standard linear channel model, the signal received by the $k$th user in a given channel use is written as
\begin{equation}
\label{Eq_yk}
y_{k} =  \mathbf{h}_{k}^{H}\mathbf{x}+n_{k}
\end{equation}
where $\mathbf{h}_{k} \in \mathbb{C}^{N_{t}}$ is the channel vector from the BS to the $k$th user,
$\mathbf{x} \in \mathbb{C}^{N_{t}}$ is the transmit signal, and $n_{k} \thicksim \mathcal{CN} ( 0 , \sigma^{2}_{\mathrm{n},{k}} )$ is the Additive White Gaussian Noise (AWGN) at the receiver.
The transmit signal is subject to an average power constraint $\mathrm{E}\{\mathbf{x}^{H}\mathbf{x}\} \leq P_{\mathrm{t}}$.
Without loss of generality, we assume equal noise variances across users, i.e. $\sigma_{\mathrm{n},{k}}^{2}=\sigma_{\mathrm{n}}^{2}$.
Therefore, the transmit SNR is defined as $\mathrm{SNR} \triangleq P_{\mathrm{t}} / \sigma_{\mathrm{n}}^{2}$.
Moreover, $\sigma_{\mathrm{n}}^{2}$ is non-zero and remains fixed over the entire SNR range. Hence, $\mathrm{SNR} \rightarrow \infty$ is equivalent to $P_{\mathrm{t}}  \rightarrow \infty$.

The BS wishes to communicate $K$ independent messages $W_{\mathrm{t},1},\ldots,W_{\mathrm{t},K}$ uniformly drawn from the sets $\mathcal{W}_{\mathrm{t},1},\ldots,\mathcal{W}_{\mathrm{t},K}$, and intended for receivers $1,\ldots,K$ respectively.
A RS scheme is considered where the message of each user is split into a private part and a common part,
i.e. $W_{\mathrm{t},k} = \{W_{k},W_{\mathrm{c},k}\}$\footnote{The subscript $\mathrm{t}$ denotes "total", which indicates that $W_{\mathrm{t},k}$ is generally composed of two parts. The subscript $\mathrm{p}$ is omitted for private parts as they resemble conventional transmission (NoRS).}
with
$W_{k} \in \mathcal{W}_{k}$, $W_{\mathrm{c},k} \in \mathcal{W}_{\mathrm{c},k}$, and
$\mathcal{W}_{k} \times \mathcal{W}_{\mathrm{c},k} = \mathcal{W}_{\mathrm{t},k}$.
A super message (known as the common message) is composed by packing the common parts such that $W_{\mathrm{c}} = \{W_{\mathrm{c},1} ,\ldots,W_{\mathrm{c},K} \} \in \mathcal{W}_{\mathrm{c},1} \times \ldots \times \mathcal{W}_{\mathrm{c},K} $.
The resulting $K+1$ messages are encoded into the independent data streams
$s_{\mathrm{c}},s_{1},\ldots,s_{K}$, where $s_{\mathrm{c}}$ and $s_{k}$
are encoded common and private symbols in an arbitrary channel use.
The $K+1$ symbols of a given channel use are grouped in a vector $\mathbf{s} \triangleq [s_{\mathrm{c}},s_{1},\ldots,s_{K}]^{T} \in \mathbb{C}^{K+1}$, where
$\mathrm{E}\{\mathbf{s}\mathbf{s}^{H}\} = \mathbf{I}$.

Symbols are mapped to the BS antennas through a linear precoding matrix defined as $\mathbf{P} \triangleq [\mathbf{p}_{\mathrm{c}},\mathbf{p}_{1},\ldots,\mathbf{p}_{K}]$, where $\mathbf{p}_{\mathrm{c}} \in \mathbb{C}^{N_{t}}$ is the common precoder, and
$\mathbf{p}_{k} \in \mathbb{C}^{N_{t}}$ is the $k$th private precoder. This yields a transmit signal which writes as
\begin{equation}\label{Eq_x}
  \mathbf{x} = \mathbf{P}\mathbf{s} = \mathbf{p}_{\mathrm{c}}s_{\mathrm{c}} + \sum_{k=1}^{K}\mathbf{p}_{k}s_{k}
\end{equation}
%
where the common signal is superimposed on top of the private signals. The power constraint is rewritten as $\mathrm{tr}\big( \mathbf{P}\mathbf{P}^{H} \big) \leq P_{\mathrm{t}}$.

The $k$th user's average receive power is written as
\begin{equation}
\label{Eq_T_c_k}
T_{\mathrm{c},k} = \overbrace{|\mathbf{h}_{k}^{H}\mathbf{p}_{\mathrm{c}}|^{2}}^{S_{\mathrm{c},k}} + \underbrace{ \overbrace{|\mathbf{h}_{k}^{H}\mathbf{p}_{k}|^{2}}^{S_{k}} + \overbrace{\sum_{i\neq k} |\mathbf{h}_{k}^{H}\mathbf{p}_{i}|^{2} + \sigma_{\mathrm{n}}^{2}}^{I_{k}}}_{I_{\mathrm{c},k} = T_{k}}.
\end{equation}
To recover its message, each receiver carries out multi-stream detection of both the common stream and its designated private stream.
The common stream is decoded first by treating all other streams as noise.
Given that the common message is successfully recovered, the common signal is removed from $y_{k}$ using SIC in order to improve the detectability
of the private stream, which is then decoded in the presence of the remaining interference.
The corresponding output SINRs of the common stream and the private stream at the $k$th receiver write as
\begin{equation}
\label{Eq_SINR}
\gamma_{\mathrm{c},k} \triangleq S_{\mathrm{c},k}I_{\mathrm{c},k}^{-1} \quad \text{and} \quad
\gamma_{k}  \triangleq S_{k}I_{k}^{-1}.
\end{equation}
Under Gaussian signalling, the corresponding achievable rates in bits per channel use are given as
\begin{equation}
\label{Eq_R}
R_{\mathrm{c},k} = \log_{2}(1+\gamma_{\mathrm{c},k})
\quad \text{and} \quad
R_{k} = \log_{2}(1+\gamma_{k}).
\end{equation}
To ensure that $W_{\mathrm{c}}$ is decodable by all users, it should be transmitted at the common rate defined as $R_{\mathrm{c}} \triangleq \min_{l}\{R_{\mathrm{c},l}\}_{l=1}^{K}$.
Following the RS structure described earlier, the common rate is split into $K$ portions, namely $C_{1},\ldots,C_{K}$ where $\sum_{k=1}^{K} C_{k} = R_{\mathrm{c}}$. Note that $C_{k}$ corresponds to the achievable rate of the common part of the $k$th user's message, i.e. $W_{\mathrm{c},k}$.
Hence, the $k$th user achieves a total rate of $R_{\mathrm{t},k} \triangleq R_{k} + C_{k}$.
\section{Problem Statement and Asymptotic Performance}
\label{Section_Problem_Statement}
For each channel vector $\mathbf{h}_{k}$, the BS obtains an erroneous estimate $\widehat{\mathbf{h}}_{k}$,
from which the error unknown to the BS is given as  $\widetilde{\mathbf{h}}_{k} \triangleq \mathbf{h}_{k} -  \widehat{\mathbf{h}}_{k}$.
As far as the BS is aware, $\widetilde{\mathbf{h}}_{k}$ is bounded by an origin-centered sphere with radius $\delta_{k}$, from which  $\mathbf{h}_{k}$ is confined within the uncertainty region
$\mathbb{H}_{k} \triangleq \left\{ \mathbf{h}_{k} \mid  \mathbf{h}_{k} = \widehat{\mathbf{h}}_{k} +   \widetilde{\mathbf{h}}_{k},
\|\widetilde{\mathbf{h}}_{k}\| \leq \delta_{k}  \right\}$.
This CSIT uncertainty model is highly relevant in limited feedback systems, where each receiver estimates its channel vector through downlink training and then sends back a quantized version to the BS \cite{Love2008,Jindal2006,Ding2007,Caire2010}.
The resulting quantization errors fall within a closed and bounded region which may either be a closed ball, or is fully contained within one.
We should highlight that we assume perfect Receiver CSI (CSIR) throughout this work. This assumption is justified as it can be shown that under proper downlink training, the effects of estimation errors are at the same power level of additive noise, and are completely overwhelmed by the influence of feedback errors \cite{Caire2010}.

%
Due to the CSIT uncertainty, the actual rates cannot be considered as design metrics at the BS.
From the BS's point of view, achievable rates also lie in bounded uncertainty regions.
We consider a robust design where precoders are optimized w.r.t the worst-case achievable rates.
The $k$th user's worst-case achievable rates are defined  as
\begin{equation}
\bar{R}_{\mathrm{c},k} \triangleq  \underset{\mathbf{h}_{k} \in \mathbb{H}_{k}}{\min}
R_{\mathrm{c},k}(\mathbf{h}_{k})
\quad \text{and} \quad
\bar{R}_{k} \triangleq  \underset{\mathbf{h}_{k} \in \mathbb{H}_{k}}{\min}
R_{k}(\mathbf{h}_{k})
\end{equation}
where the dependencies of the rates on $\mathbf{h}_{k}$ are highlighted. The worst-case achievable common rate is defined as
$\bar{R}_{\mathrm{c}} \triangleq \min_{l}\{\bar{R}_{\mathrm{c},l}\}_{l=1}^{K}$.
Transmitting $W_{\mathrm{c}},W_{1},\ldots,W_{K}$ at rates $\bar{R}_{\mathrm{c}},\bar{R}_{1},\ldots,\bar{R}_{K}$ respectively, guarantees successful decoding at the receivers for all admissible channels within the uncertainty regions.
Given some splitting ratios, the $k$th user's portion of the worst-case common rate is denoted by $\bar{C}_{k}$ where $\sum_{k=1}^{K}\bar{C}_{k} = \bar{R}_{\mathrm{c}}$.
Therefore, the $k$th user's worst-case total achievable rate is given as $\bar{R}_{\mathrm{t},k} \triangleq \bar{R}_{k} + \bar{C}_{k}$, corresponding to the worst-case transmission rate of the original message $W_{\mathrm{t},k}$.
%
\subsection{Max-Min Fairness Optimization Problem}
The robust optimization problem of achieving max-min fairness using the RS transmission strategy is posed as
%
\begin{equation}
\label{Eq_Problem_R_RS}
\mathcal{R}_{\mathrm{RS}}(P_{\mathrm{t}}):
\begin{cases}
       \underset{ \bar{R}_{\mathrm{t}},\bar{\mathbf{c}},\mathbf{P} }{\max} & \bar{R}_{\mathrm{t}} \\
       \text{s.t.}  & \bar{R}_{k} + \bar{C}_{k}  \geq  \bar{R}_{\mathrm{t}},\; \forall k \in  \mathcal{K} \\
                    & \bar{R}_{\mathrm{c},k} \geq \sum_{l=1}^{K} \bar{C}_{l}, \; \forall k\in\mathcal{K} \\
                    & \bar{C}_{k} \geq 0, \; \forall k \in \mathcal{K}\\
                    & \mathrm{tr}\big(\mathbf{P}\mathbf{P}^{H}\big) \leq P_{\mathrm{t}}
\end{cases}
\end{equation}
where $\bar{R}_{\mathrm{t}}$ is an auxiliary variable and $\bar{\mathbf{c}} \triangleq [\bar{C}_{1},\ldots,\bar{C}_{K}]^{T}$.
The pointwise minimizations in the original objective function is replaced with the inequality constraints in \eqref{Eq_Problem_R_RS}.
For example, $\bar{R}_{k} + \bar{C}_{k}  \geq  \bar{R}_{\mathrm{t}}$ guarantees fairness, while
$\bar{R}_{\mathrm{c},k} \geq \sum_{l=1}^{K} \bar{C}_{l}$ guarantees that the common message is decoded by the $k$th user.
The constraint $\bar{C}_{k} \geq 0$ however is to guarantee the non-negativity of the common rate portions.

In contrast, the NoRS version of the problem is give by
\begin{equation}
\label{Eq_Problem_R_NoRS}
\mathcal{R}(P_{\mathrm{t}}):
\begin{cases}
       \underset{\bar{R},\mathbf{P}_{\mathrm{p}}}{\max} & \bar{R} \\
       \text{s.t.}  & \bar{R}_{k} \geq \bar{R}, \;  \forall k\in\mathcal{K} \\
                    & \mathrm{tr}\big(\mathbf{P}_{\mathrm{p}}\mathbf{P}_{\mathrm{p}}^{H}\big) \leq P_{\mathrm{t}}.
\end{cases}
\end{equation}
where $\bar{R}$ is the rate auxiliary variable, and $\mathbf{P}_{\mathrm{p}}\triangleq [\mathbf{p}_{1},\ldots,\mathbf{p}_{K}]$.
It is evident that solving \eqref{Eq_Problem_R_NoRS} is equivalent to solving \eqref{Eq_Problem_R_RS} over a restricted domain characterized by setting $\bar{\mathbf{c}} = \mathbf{0}$, which in turn forces
$\|\mathbf{p}_{\mathrm{c}}\|^{2}$ to zero at optimality. As a result, we can write $\mathcal{R}_{\mathrm{RS}}(P_{\mathrm{t}}) \geq \mathcal{R}(P_{\mathrm{t}})$.
Before proceeding to derive the optimum performances, we adapt the concept of CSIT error scaling in \cite{Jindal2006,Yang2013,Hao2013,Hao2015,Davoodi2014} to the bounded error model in this work.
\subsection{CSIT Error Scaling}
It is well established that CSIT uncertainties hinder the performance of MU-MISO systems, as it may be impossible to design precoding schemes that eliminate the received MU interference or reduce it to the level of noise.
Taking the considered model as an example, the SINR of the private stream given in \eqref{Eq_SINR}
will have residual interference terms of power
$\sum_{i \neq k} \mid \widetilde{\mathbf{h}}_{k}^{H} \mathbf{p}_{i} \mid^{2}$ (at least), scaling as
$\delta_{k}^{2} \sum_{i \neq k} \|\mathbf{p}_{i} \|^{2}$ in the worst-case sense (see Appendix \ref{Appendix_Proof_Theorem_RS_NoRS_Max_Min_DoF}).
For the NoRS problem in \eqref{Eq_Problem_R_NoRS}, increasing the total transmit power $P_{\mathrm{t}}$ (or equivalently the SNR) should be accompanied by a simultaneous increase in all user powers, allocated solely to private streams in this case, as otherwise users with fixed powers will experience a degree of unfairness.
Under fixed $\delta_{1}^{2},\ldots,\delta_{K}^{2}$ which are independent of $P_{\mathrm{t}}$, residual interference becomes dominant as
$\|\mathbf{p}_{1} \|^{2}, \ldots,\|\mathbf{p}_{K} \|^{2}$ grow high w.r.t the noise level, yielding a saturating rate performance at high SNRs.
This phenomenon was noted by \emph{Jindal} in \cite{Jindal2006}, where Zero Forcing Beamforming (ZF-BF) under a Random Vector Quantization (RVQ) type of feedback was considered. \emph{Jindal} suggested improving the CSIT quality with the SNR level through increasing the number of feedback bits.
The intuition behind this feedback scaling is simply explained as follows: in the high SNR regime where performance is interference limited,  interference nulling is crucial to guarantee non-saturating performance, which necessitates higher CSIT accuracy, or equivalently, decaying CSIT errors.
Incorporating this concept into the considered error model yields uncertainty regions that shrink with increased SNR given that the number of feedback bits is scaled accordingly.
Equivalently, we write
$\delta_{k}^{2} = O(P_{\mathrm{t}}^{-\alpha_{k}})$, where  $\alpha_{k} \in [0,\infty)$ is a constant exponent that quantifies the CSIT quality
(or scaling law) as SNR grows large, i.e.
$\alpha_{k} \triangleq \lim_{P_{\mathrm{t}}\rightarrow\infty} -\frac{\log(\delta_{k}^{2})}{\log(P_{\mathrm{t}})}$.
$\alpha_{k} =0$ represents a constant (or very slowly scaling) number of feedback bits, yielding a non-shrinking (fixed) uncertainty region.
On the other hand,
$\alpha_{k} = \infty$ corresponds to perfect CSIT resulting from an infinitely high number of feedback bits.
In the following, the exponents are truncated such that $\alpha_{k} \in [0,1]$.
This is customary in asymptotic rate analysis as $\alpha_{k} = 1$ corresponds to perfect CSIT in the DoF sense \cite{Jindal2006,Yang2013,Davoodi2014}.
This can be shown by noting that the worst-case residual interference would scale as $O(P_{\mathrm{t}}^{0})$ at most, and hence has a similar effect to additive noise. Without loss of generality, we assume that the qualities are ordered as $\alpha_{1} \leq \alpha_{2} \leq \ldots \leq \alpha_{K}$.
\subsection{DoF Analysis}
Due to the crucial role of MU interference and its hindering influence in the presence of CSIT uncertainty, the performance is characterized in terms of the DoF: a first order approximation of the achievable rate in the high SNR regime, roughly
interpreted as the number of interference-free streams that can be simultaneously communicated in a single channel use.
To facilitate this analysis, we define a precoding scheme for \eqref{Eq_Problem_R_RS} as a family of feasible precoders with
one precoding matrix for each SNR level,
i.e. $\big\{ \mathbf{P}(P_{\mathrm{t}}) \big\}_{P_{\mathrm{t}}}$.
The associated powers allocated to the precoding vectors are defined as $q_{\mathrm{c}} \triangleq \| \mathbf{p}_{\mathrm{c}} \|^{2}$ and $q_{k} \triangleq \| \mathbf{p}_{k} \|^{2}$, which are assumed to scale as $O(P_{\mathrm{t}}^{a_{\mathrm{c}}})$ and $O(P_{\mathrm{t}}^{a_{k}})$ respectively, where
$a_{\mathrm{c}},a_{k} \in [0,1]$ are some scaling exponents.
For a given precoding scheme, the worst-case achievable DoFs write as
\begin{equation}
\label{Eq_Wors_Case_DoF}
\bar{d}_{\mathrm{c}}
\triangleq \lim_{P_{\mathrm{t}} \rightarrow \infty} \frac{\bar{R}_{\mathrm{c}}(P_{\mathrm{t}})}{\log_{2}(P_{\mathrm{t}})}
\quad \text{and} \quad
\bar{d}_{k}
\triangleq \lim_{P_{\mathrm{t}} \rightarrow \infty} \frac{\bar{R}_{k}(P_{\mathrm{t}})}{\log_{2}(P_{\mathrm{t}})}
\end{equation}
where dependencies on the power level are highlighted in \eqref{Eq_Wors_Case_DoF}.
The portion of $\bar{d}_{\mathrm{c}}$ allocated to the  $k$th user is given by
$\bar{c}_{k} \triangleq \lim_{P_{\mathrm{t}} \rightarrow \infty} \frac{\bar{C}_{k}(P_{\mathrm{t}})}{\log_{2}(P_{\mathrm{t}})}$,
where $\sum_{k=1}^{K} \bar{c}_{k} = \bar{d}_{\mathrm{c}}$.
In the following, the worst-case DoF is simply referred to as the DoF for brevity.
All definitions extend to the NoRS case where the common part is discarded, and a precoding scheme for \eqref{Eq_Problem_R_NoRS} writes as $\big\{ \mathbf{P}_{\mathrm{p}}(P_{\mathrm{t}}) \big\}_{P_{\mathrm{t}}}$.
As each receiver is equipped with a single antenna, it can decode one interference free stream at most, from which we have $\bar{d}_{\mathrm{c}} \leq 1$, $\bar{d}_{k} \leq 1$ and $\bar{d}_{\mathrm{c}} + \bar{d}_{k} \leq 1$.
The optimum max-min DoF is derived under the following assumption regarding the condition of the channel.
%
\newtheorem{Assumption_Bounded_Channel}[Assumption_Counter]{Assumption}
\begin{Assumption_Bounded_Channel}
\label{Assumption_Bounded_Channel}
\textnormal{ Although the channel estimates and errors may depend on SNR, the actual channel vectors $\mathbf{h}_{1},\ldots,\mathbf{h}_{K}$ do not, and their entries are assumed to have absolute values bounded away from zero and infinity.
Moreover, the channel estimate matrix $\widehat{\mathbf{H}} \triangleq \big[ \widehat{\mathbf{h}}_{1},\ldots,\widehat{\mathbf{h}}_{K} \big]$ is of full column rank.
}
\end{Assumption_Bounded_Channel}
When the entries of the actual channel are drawn from continuous unbounded distributions, the assumption can be made by reducing the probability measure of the omitted neighborhoods to an arbitrary small value with no impact on the DoF \cite{Davoodi2014}.
On the other hand, the second part can be guaranteed in a feedback system by prohibiting the scheduling of users with similar quantized channel vectors in the same time-frequency slot. However, it should be noted that while the second part of Assumption \ref{Assumption_Bounded_Channel} is required for the proof of the following result, it is not necessary for the optimization solutions in the following sections, as problems \eqref{Eq_Problem_R_RS} and \eqref{Eq_Problem_R_NoRS} can still be solved under linearly dependent channel estimates.
%
\newtheorem{Theorem_RS_NoRS_Max_Min_DoF}[Theorem_Counter]{Theorem}
\begin{Theorem_RS_NoRS_Max_Min_DoF}\label{Theorem_RS_NoRS_Max_Min_DoF}
\textnormal{
For the NoRS problem defined in \eqref{Eq_Problem_R_NoRS}, the optimum max-min DoF is given by
\begin{equation}
\label{Eq_Max_Min_DoF_NoRS}
\bar{d}^{\ast} \triangleq \lim_{P_{\mathrm{t}} \rightarrow \infty} \frac{\mathcal{R}(P_{\mathrm{t}})}{\log(P_{\mathrm{t}})} = \frac{\alpha_{1}+\alpha_{2}}{2}.
\end{equation}
The RS problem in \eqref{Eq_Problem_R_RS} yields an optimum max-min DoF of
\begin{equation}
\label{Eq_Max_Min_DoF_RS}
\bar{d}^{\ast}_{\mathrm{RS}} \triangleq \lim_{P_{\mathrm{t}} \rightarrow \infty} \frac{\mathcal{R}_{\mathrm{RS}}(P_{\mathrm{t}})}{\log(P_{\mathrm{t}})} = \min_{J \in \{2,\ldots,K\}} \frac{1+\sum_{k=1}^{J-1} \alpha_{k}}{J}.
\end{equation}
}
\end{Theorem_RS_NoRS_Max_Min_DoF}
The results are obtained through two steps. First, we show that the max-min DoFs are upper-bounded by \eqref{Eq_Max_Min_DoF_NoRS} and \eqref{Eq_Max_Min_DoF_RS},
then we show that the upper-bounds are achievable via feasible precoding schemes.
The full proof is relegated to Appendix \ref{Appendix_Proof_Theorem_RS_NoRS_Max_Min_DoF}.
In the following discussion, we provide some insight into Theorem \ref{Theorem_RS_NoRS_Max_Min_DoF} and its implications.
\eqref{Eq_Max_Min_DoF_NoRS} shows that the NoRS optimum max-min DoF is determined by the worst two CSIT scaling factors.
This is explained through the following example.
Consider that all receivers are switched off except for user-1 and user-2.
A simultaneous and proportional increase in powers allocated to the two users combined with a proper design of precoders achieves $\bar{d}_{1} = \alpha_{1}$ and $\bar{d}_{2} = \alpha_{2}$ (see Lemma \ref{Lemma_achievable_DoF}).
Compromising user-2's DoF by reducing its power scaling reduces the interference experienced by user-1, and both users can achieve the DoF in \eqref{Eq_Max_Min_DoF_NoRS}.
Moving back to the $K$-user case by introducing users with possibly better CSIT qualities does not improve the max-min DoF.
In particular, even if the new users can achieve higher DoF (ultimately 1), the max-min DoF is limited by that of user-1 and user-2, i.e. $\frac{\alpha_{1}+\alpha_{2}}{2}$.
Hence, the best they could do is achieve an equal (or higher) DoF without influencing $\bar{d}_{1}$ and $\bar{d}_{2}$, which can be shown to be possible given their better CSIT qualities.

Considering the same 2-user example under RS, allocating private powers that scale as $O\big(P_{\mathrm{t}}^{\alpha_{1}}\big)$ and a common power that scales as $O\big(P_{\mathrm{t}}\big)$ yields $\bar{d}_{1},\bar{d}_{2} =  \alpha_{1}$ and $\bar{d}_{\mathrm{c}} = 1 - \alpha_{1}$ as shown in Lemma \ref{Lemma_achievable_DoF}.
By Splitting $\bar{d}_{\mathrm{c}}$ evenly, each user obtains a total DoF of $\frac{1+\alpha_{1}}{2}$.
Increasing $K$ may decrease this max-min DoF as $\bar{d}_{\mathrm{c}}$ may be divided among a larger set of users, as reflected by the minimization  in \eqref{Eq_Max_Min_DoF_RS}.
However, $\bar{d}_{\mathrm{RS}}^{\ast} > \bar{d}^{\ast} $ holds for all $ \alpha_{1},\ldots,\alpha_{K} \in [0,1)$,
i.e. RS provides a strict improvement over NoRS under imperfect CSIT in a DoF sense.
Note that the existence of two or more users with $\alpha_{k} = 0$ yields $\bar{d}^{\ast} = 0$.
This results in a saturating max-min rate performance for NoRS as observed in \cite{Jindal2006,Shenouda2007}.
On the contrary, $\bar{d}^{\ast}_{\mathrm{RS}}$ is lower-bounded by $\bar{d}^{\ast}_{\mathrm{RS}} \geq 1/K$ regardless of the CSIT scaling, and hence achieves an ever-growing max-min rate.
In conclusion of this section, we highlight that while rate optimality implies DoF optimality, the converse does not hold in general. In particular, the optimum DoF performance characterized in Theorem \ref{Theorem_RS_NoRS_Max_Min_DoF} can be achieved via suboptimal precoders (in the rate sense) as seen in Appendix \ref{Appendix_Proof_Theorem_RS_NoRS_Max_Min_DoF}. This common observation has motivated a number of suboptimal designs in the MIMO literature, e.g. the employment of the rate-suboptimal yet DoF-optimum ZF-BF strategy \cite{Caire2003,Jindal2006,Davoodi2014}.
The implication of this remark is witnessed in Section \ref{Section_Simulation_Results},
where simulation results show that the (possibly) suboptimal precoders designed in Section \ref{Section_Cutting_Set} achieve the optimum DoF.
\section{Conservative Approach}
\label{Section_Conservative}
The optimization problem in \eqref{Eq_Problem_R_RS} is non-convex and semi-infinite.
Even a sampled instance of the problem with a finite number of constraints is highly intractable in its current form due to the non-convex coupled nature of the sum-rate expressions embedded in each user's achievable rate.
We start this section by introducing the WMMSE approach, initially proposed in \cite{Christensen2008} and then further established in \cite{Shi2011,Razaviyayn2013b} with more rigorous convergence proofs.
This approach is heavily employed throughout this paper due to its effectiveness in solving problems featuring sum-rate expressions.
\subsection{Rate-WMMSE Relationship}
\label{Subsection_Rate_WMSE}
Let $\widehat{s}_{\mathrm{c},k}=g_{\mathrm{c},k} y_{k}$ be the $k$th user's estimate of $s_{\mathrm{c}}$ obtained by applying the scalar equalizer $g_{\mathrm{c},k}$.
Given that the common message is successfully decoded and removed from $y_{k}$, the estimate of $s_{k}$ is obtained by applying $g_{k}$ such that $\widehat{s}_{k}=g_{k}(y_{k}-\mathbf{h}_{k}^{H}\mathbf{p}_{\mathrm{c}}s_{\mathrm{c}})$.
The corresponding common and private MSEs, defined as $\varepsilon_{\mathrm{c},k} \triangleq \mathrm{E}\{|\widehat{s}_{\mathrm{c},k} - s_{\mathrm{c}}|^{2}\}$ and $\varepsilon_{k} \triangleq \mathrm{E}\{|\widehat{s}_{k} - s_{k}|^{2}\}$ respectively, write as:
\begin{subequations}
\label{Eq_MSE}
\begin{align}
  \label{Eq_MSE_c_k}
  \varepsilon_{\mathrm{c},k} & = |g_{\mathrm{c},k}|^{2} T_{\mathrm{c},k} -2\Re \big\{g_{\mathrm{c},k}\mathbf{h}_{k}^{H}\mathbf{p}_{\mathrm{c}}\big\}+1 \\
  \label{Eq_MSE_k}
  \varepsilon_{k} & =  |g_{k}|^{2} T_{k}-2\Re \big\{g_{k}\mathbf{h}_{k}^{H}\mathbf{p}_{k}\big\}+1
\end{align}
\end{subequations}
where $T_{k}$ is defined in \eqref{Eq_T_c_k}.
Optimum equalizers are obtained from $\frac{\partial \varepsilon_{\mathrm{c},k}}{\partial g_{\mathrm{c},k}} = 0$ and
$\frac{\partial \varepsilon_{k}}{\partial g_{k}} = 0$. This yields the well-known MMSE
equalizers given by
\begin{equation}
 \label{Eq_g_MMSE}
  g_{\mathrm{c},k}^{\mathrm{MMSE}} = \mathbf{p}_{\mathrm{c}}^{H}\mathbf{h}_{k} T_{\mathrm{c},k}^{-1}
  \quad \text{and} \quad
  g_{k}^{\mathrm{MMSE}} = \mathbf{p}_{k}^{H}\mathbf{h}_{k}T_{k}^{-1}
\end{equation}
from which the  MMSEs are obtained as
\begin{subequations}
\label{Eq_MMSE}
\begin{align}
\label{Eq_MMSE_c}
\varepsilon_{\mathrm{c},k}^{\mathrm{MMSE}} & \triangleq \underset{g_{\mathrm{c},k}}{\min} \ \varepsilon_{\mathrm{c},k} =  T_{\mathrm{c},k}^{-1} I_{\mathrm{c},k} \\
\label{Eq_MMSE_k}
\varepsilon_{k}^{\mathrm{MMSE}} & \triangleq \underset{g_{k}}{\min} \ \varepsilon_{k} = T_{k}^{-1}I_{k}.
\end{align}
\end{subequations}
Under optimum equalization, the SINRs in \eqref{Eq_SINR} relate to the MMSEs such that
$\gamma_{\mathrm{c},k} = \big( 1/\varepsilon_{\mathrm{c},k}^{\mathrm{MMSE}} \big) - 1 $ and
$\gamma_{k} = \big( 1/\varepsilon_{k}^{\mathrm{MMSE}} \big) - 1 $,
from which the rates in \eqref{Eq_R} write as $R_{\mathrm{c},k} =-\log_{2}(\varepsilon_{\mathrm{c},k}^{\mathrm{MMSE}})$
and $R_{k} =-\log_{2}(\varepsilon_{k}^{\mathrm{MMSE}})$.

By introducing positive weights  $(u_{\mathrm{c},k}, u_{k})$, the $k$th user's common and private augmented WMSEs are defined as:
\begin{equation}
\label{Eq_A_WMSEs}
\xi_{\mathrm{c},k}
 \triangleq
u_{\mathrm{c},k} \varepsilon_{\mathrm{c},k}  -  \log_{2} (  u_{\mathrm{c},k} )
\ \  \text{and}  \ \
\xi_{k}
 \triangleq
u_{k} \varepsilon_{k}  -  \log_{2}  (  u_{k} ).
\end{equation}
In the following, $\xi_{\mathrm{c},k}$ and $\xi_{k}$ are referred to as  the WMSEs for brevity.
Optimizing over the equalizers and weights, the Rate-WMMSE relationship writes as:
\begin{subequations}
\label{Eq_min_WMSE}
\begin{align}
 \xi_{\mathrm{c},k}^{\mathrm{MMSE}} & \triangleq  \underset{u_{\mathrm{c},k}, g_{\mathrm{c},k}}{\min} \xi_{\mathrm{c},k} = 1-R_{\mathrm{c},k}
\\
 \xi_{k}^{\mathrm{MMSE}} & \triangleq \underset{u_{k}, g_{k}}{\min} \ \xi_{k}= 1-R_{k}
\end{align}
\end{subequations}
where the optimum equalizers are given by: $g_{\mathrm{c},k}^{\ast}  = g_{\mathrm{c},k}^{\mathrm{MMSE}} $ and $g_{k}^{\ast} = g_{k}^{\mathrm{MMSE}}$, and the optimum weights are given by:
$u_{\mathrm{c},k}^{\ast} = u_{\mathrm{c},k}^{\mathrm{MMSE}} \triangleq \big( \varepsilon_{\mathrm{c},k}^{\mathrm{MMSE}} \big)^{-1}$
and
$u_{k}^{\ast} = u_{k}^{\mathrm{MMSE}} \triangleq \big( \varepsilon_{k}^{\mathrm{MMSE}} \big)^{-1}$,
obtained by checking the first order optimality conditions.
$ \xi_{\mathrm{c},k}$ (resp. $ \xi_{k}$) is convex in each of $g_{\mathrm{c},k}$ (resp. $g_{k}$),
$u_{\mathrm{c},k}$ (resp. $u_{k}$), and $\mathbf{P}$, while fixing the two other variables.
This variable-wise convexity, preserved under sum-WMSE expressions, alongside the relationship in \eqref{Eq_min_WMSE} are key to solving non-convex optimization problems with sum-rate expressions.
By incorporating the CSIT uncertainty into \eqref{Eq_min_WMSE}, the worst-case rates can be equivalently expressed as
\begin{subequations}
\label{Eq_wc_Rates_WMSEs}
\begin{align}
\bar{R}_{\mathrm{c},k} & = 1 - \max_{\mathbf{h}_{k} \in \mathbb{H}_{k}} \min_{u_{\mathrm{c},k}, g_{\mathrm{c},k}}
\xi_{\mathrm{c},k}\big( \mathbf{h}_{k}, g_{\mathrm{c},k}, u_{\mathrm{c},k} \big) \\
\bar{R}_{k} & = 1 - \max_{\mathbf{h}_{k} \in \mathbb{H}_{k}} \min_{u_{k}, g_{k}}
\xi_{k}\big( \mathbf{h}_{k}, g_{k}, u_{k} \big).
\end{align}
\end{subequations}
Plugging \eqref{Eq_wc_Rates_WMSEs}'s right-hand side expressions into \eqref{Eq_Problem_R_RS} yields an equivalent WMSE problem with an extended domain which includes the equalizers and weights as optimization variables.
The equivalent problem inherits the virtue of WMSEs, harboring a block-wise convex structure, i.e. convex in each block of variables while fixing the rest.
Such structure can be exploited to obtain a solution using the AO principle, also known as the Block Coordinate Descent (BCD) method \cite{Bertsekas1999}.
However, despite its desirable properties, the WMSE problem is still deemed intractable due to its infinitely many variables and constraints.
This follows from the dependencies of the optimum equalizers and weights on perfect CSI, which is seen from \eqref{Eq_min_WMSE} and noting that the maximizations are outside the minimizations in \eqref{Eq_wc_Rates_WMSEs}.
This hurdle is addressed through a conservative approximation in this section, and a sampling-based method in the following section.
%
\subsection{Conservative Worst-Case Approximation}
%
The infinite sets of variables are abstracted into finite sets by leveraging the conservative approximation in \cite{Tajer2011}. We write
\begin{subequations}
\label{Eq_wc_Rates_WMSEs_LB}
\begin{align}
\label{Eq_wc_Rates_WMSEs_c_LB}
\widehat{R}_{\mathrm{c},k} & = 1 -  \min_{\widehat{u}_{\mathrm{c},k}, \widehat{g}_{\mathrm{c},k}} \max_{\mathbf{h}_{k} \in \mathbb{H}_{k}}
\xi_{\mathrm{c},k}\big( \mathbf{h}_{k}, \widehat{g}_{\mathrm{c},k}, \widehat{u}_{\mathrm{c},k} \big) \\
\widehat{R}_{k} & = 1 -  \min_{\widehat{u}_{k}, \widehat{g}_{k}} \max_{\mathbf{h}_{k} \in \mathbb{H}_{k}}
\xi_{k}\big( \mathbf{h}_{k}, \widehat{g}_{k}, \widehat{u}_{k} \big)
\end{align}
\end{subequations}
where $\widehat{R}_{\mathrm{c},k} \leq \bar{R}_{\mathrm{c},k}$ and
$\widehat{R}_{k} \leq \bar{R}_{k}$ are lower-bounds on the worst-case rates (see footnote 1 in \cite[Section IV.B.2]{Tajer2011}), and $(\widehat{g}_{\mathrm{c},k},\widehat{g}_{k})$ and $(\widehat{u}_{\mathrm{c},k},\widehat{u}_{k})$ are the abstracted equalizers and weights.
By swapping the order of the minimization (optimization) and maximization (worst-case channel) as shown in \eqref{Eq_wc_Rates_WMSEs_LB}, the equalizers and weights loose their dependencies on perfect CSI.
Taking \eqref{Eq_wc_Rates_WMSEs_c_LB} for example, the same $(\widehat{u}_{\mathrm{c},k},\widehat{g}_{\mathrm{c},k})$ are employed for all realizations $\mathbf{h}_{k} \in \mathbb{H}_{k}$.
Plugging \eqref{Eq_wc_Rates_WMSEs_LB}  into problem \eqref{Eq_Problem_R_RS} yields the conservative WMSE counterpart
%
\begin{equation}
\label{Eq_Problem_WMSE_R_RS_LB}
\widehat{\mathcal{R}}_{\mathrm{RS}}(P_{\mathrm{t}}) \! : \!
\begin{cases}
       \underset{ \widehat{R}_{\mathrm{t}},\widehat{\mathbf{c}},\mathbf{P},\widehat{\mathbf{g}},\widehat{\mathbf{u}}}
       {\max}
       \ \ \widehat{R}_{\mathrm{t}} \\
       \text{s.t.} \\
       1 \! - \! \xi_{k}\big( \mathbf{h}_{k}, \widehat{g}_{k}, \widehat{u}_{k} \big) \! + \! \widehat{C}_{k}  \geq  \widehat{R}_{\mathrm{t}}, \forall \mathbf{h}_{k} \in \mathbb{H}_{k}, k \in  \mathcal{K} \\
       1 - \xi_{\mathrm{c},k}\big( \mathbf{h}_{k}, \widehat{g}_{\mathrm{c},k}, \widehat{u}_{\mathrm{c},k} \big)  \geq
       \sum_{l=1}^{K} \widehat{C}_{l}, \forall \mathbf{h}_{k} \in \mathbb{H}_{k}, \\
        k \in  \mathcal{K} \\
          \widehat{C}_{k} \geq 0, \; \forall k \in \mathcal{K}\\
           \mathrm{tr}\big(\mathbf{P}\mathbf{P}^{H}\big) \leq P_{\mathrm{t}}
\end{cases}
\end{equation}
where $\widehat{\mathbf{g}} \triangleq \{\widehat{g}_{\mathrm{c},k},\widehat{g}_{k} \mid k \in \mathcal{K}\} $
and $\widehat{\mathbf{u}} \triangleq \{\widehat{u}_{\mathrm{c},k},\widehat{u}_{k} \mid k \in \mathcal{K}\} $.
Since the WMSE constraints in \eqref{Eq_Problem_WMSE_R_RS_LB} are decoupled in their equalizer-weight pairs, each pair can be optimized separately as formulated in \eqref{Eq_wc_Rates_WMSEs_LB}.
This is shown by noting that for an instance of problem \eqref{Eq_Problem_WMSE_R_RS_LB} with a given $\mathbf{P}$, the optimum solution of \eqref{Eq_wc_Rates_WMSEs_LB} maximizes all left-hand sides of the common rate inequalities.
This in turn yields a maximized $\sum_{l=1}^{K} \widehat{C}_{l}$ on the right-hand side of the inequalities.
A similar argument is made for private pairs and constraints, yielding a maximized objective $\widehat{R}_{\mathrm{t}}$.
The semi-infiniteness in \eqref{Eq_Problem_WMSE_R_RS_LB} on the other hand is eliminated by reformulating the infinite sets of constraints into equivalent finite constraints with Linear Matrix Inequalities (LMIs) using the result in \cite{Eldar2004}, based on the $\mathcal{S}$-lemma.
The $k$th user's private rate constraints in \eqref{Eq_Problem_WMSE_R_RS_LB} are rewritten as
\begin{subequations}
\label{Eq_finite_WMSE_constraints_k}
\begin{align}
\widehat{u}_{k} \big( \tau_{k} +  |\widehat{g}_{k}|^{2}\sigma_{\mathrm{n}}^{2} \big) - \log_{2}(\widehat{u}_{k}) \leq \ &
1+\widehat{C}_{k} - \widehat{R}_{\mathrm{t}} \\
\label{Eq_finite_WMSE_constraints_k_2}
\left[
  \begin{array}{ccc}
    \tau_{k} - \lambda_{k} & \bm{\psi}_{k}^{H}  & \mathbf{0}^{T}  \\
     \bm{\psi}_{k}  &  \mathbf{I}  &  -\delta_{k}\mathbf{P}_{\mathrm{p}}^{H}\widehat{g}_{k}^{H} \\
     \mathbf{0}     &-\delta_{k}\widehat{g}_{k}\mathbf{P}_{\mathrm{p}} & \lambda_{k} \mathbf{I}  \\
  \end{array}
\right] & \succeq 0
\end{align}
\end{subequations}
where $\bm{\psi}_{k}^{H} \triangleq \widehat{g}_{k}\widehat{\mathbf{h}}_{k}^{H} \mathbf{P}_{\mathrm{p}} - \mathbf{e}_{k}^{T}$.
Similarly, the common rate constraints are expressed as
\begin{subequations}
\label{Eq_finite_WMSE_constraints_c}
\begin{align}
\widehat{u}_{\mathrm{c},k} \big(\tau_{\mathrm{c},k} +  |\widehat{g}_{\mathrm{c},k}|^{2}\sigma_{\mathrm{n}}^{2} \big) - \log_{2}(\widehat{u}_{\mathrm{c},k}) \leq \ &
1 - \widehat{R}_{\mathrm{c}} \\
\label{Eq_finite_WMSE_constraints_c_2}
\left[ \! \!
  \begin{array}{ccc}
    \tau_{\mathrm{c},k} - \lambda_{\mathrm{c},k} & \bm{\psi}_{\mathrm{c},k}^{H}  & \mathbf{0}^{T}  \\
     \bm{\psi}_{\mathrm{c},k}  &  \mathbf{I}  &  -\delta_{k}\mathbf{P}^{H}\widehat{g}_{\mathrm{c},k}^{H} \\
     \mathbf{0}    &-\delta_{k}\widehat{g}_{\mathrm{c},k}\mathbf{P} & \lambda_{\mathrm{c},k} \mathbf{I}  \\
  \end{array}
  \! \!
\right]     \succeq     0
\end{align}
\end{subequations}
where $\bm{\psi}_{\mathrm{c},k}^{H} \triangleq \widehat{g}_{\mathrm{c},k}\widehat{\mathbf{h}}_{k}^{H} \mathbf{P} - \mathbf{e}_{1}^{T}$.
This is obtained by writing the common and private MSEs as
$\parallel \widehat{g}_{\mathrm{c},k} \mathbf{h}_{k}^{H} \mathbf{P} - \mathbf{e}_{1}^{T} \parallel^{2} + |\widehat{g}_{\mathrm{c},k}|^{2}\sigma_{\mathrm{n}}^{2} $
and
$\parallel \widehat{g}_{k} \mathbf{h}_{k}^{H} \mathbf{P}_{\mathrm{p}} - \mathbf{e}_{k}^{T} \parallel^{2} + |\widehat{g}_{k}|^{2}\sigma_{\mathrm{n}}^{2}$
respectively,
followed by applying the Schur Complement \cite{Horn2012}, and then the result in \cite[Proposition 2]{Eldar2004}.
For a more detailed description of the procedure, readers are referred to \cite{Vucic2009a} and \cite{Tajer2011}.
\subsection{Alternating Optimization Algorithm}
\label{Subsection_AO_conservative}
By invoking the inequalities in \eqref{Eq_finite_WMSE_constraints_k} and \eqref{Eq_finite_WMSE_constraints_c},
problem \eqref{Eq_Problem_WMSE_R_RS_LB} is solved using the AO principle.
First, $\widehat{\mathbf{g}}$ is optimized by solving the problems
$\underset{\widehat{g}_{\mathrm{c},k}}{\min} \underset{\mathbf{h}_{k} \in \mathbb{H}_{k}}{\max}
\varepsilon_{\mathrm{c},k}\big( \mathbf{h}_{k}, \widehat{g}_{\mathrm{c},k} \big)$
and
$\underset{\widehat{g}_{k}}{\min} \underset{\mathbf{h}_{k} \in \mathbb{H}_{k}}{\max}
\varepsilon_{k}\big( \mathbf{h}_{k}, \widehat{g}_{k} \big)$ for all $k\in \mathcal{K}$,
written with objective functions
$\tau_{\mathrm{c},k} +  |\widehat{g}_{\mathrm{c},k}|^{2}\sigma_{\mathrm{n}}^{2}$
and
$\tau_{k} +  |\widehat{g}_{k}|^{2}\sigma_{\mathrm{n}}^{2}$ respectively,
and constraints given in \eqref{Eq_finite_WMSE_constraints_c_2}  and \eqref{Eq_finite_WMSE_constraints_k_2} respectively\footnote{Reduced to minimizing the worst-case MSEs as weights are fixed.}.
Such problems are posed as Semidefinite Programs (SDPs)\footnote{$|\widehat{g}_{\mathrm{c},k}|^{2}$ and $|\widehat{g}_{k}|^{2}$ can be expressed as LMIs using the Schur Complement.}
and can be solved efficiently using interior-point methods \cite{Boyd2004}.
The resulting conservative MMSEs are denoted by $\widehat{\varepsilon}_{k}^{\mathrm{MMSE}}$ and $\widehat{\varepsilon}_{\mathrm{c},k}^{\mathrm{MMSE}}$.
Next, the weights are updated as $\widehat{u}_{k} = 1/\widehat{\varepsilon}_{k}^{\mathrm{MMSE}}$ and $\widehat{u}_{\mathrm{c},k} = 1/\widehat{\varepsilon}_{\mathrm{c},k}^{\mathrm{MMSE}}$.
Finally, $(\widehat{R}_{\mathrm{t}},\widehat{\mathbf{c}},\mathbf{P})$ are updated by solving problem $\widehat{\mathcal{R}}_{\mathrm{RS}}^{\mathrm{MMSE}}(P_{\mathrm{t}},\widehat{\mathbf{g}},\widehat{\mathbf{u}})$,
formulated by plugging both \eqref{Eq_finite_WMSE_constraints_k} and \eqref{Eq_finite_WMSE_constraints_c} into \eqref{Eq_Problem_WMSE_R_RS_LB} and fixing $(\widehat{\mathbf{g}},\widehat{\mathbf{u}})$.
$\widehat{\mathcal{R}}_{\mathrm{RS}}^{\mathrm{MMSE}}(P_{\mathrm{t}},\widehat{\mathbf{g}},\widehat{\mathbf{u}})$  is also a SDP, solved by interior-point methods.
This procedure is repeated in an iterative manner as described in Algorithm \ref{Algthm_AO_Conservative}.
Algorithm \ref{Algthm_AO_Conservative} is guaranteed to converge since the objective function is bounded for a given power constraint, and increases monotonically at each iteration.
The conservative approximations guarantee that the solution obtained from Algorithm \ref{Algthm_AO_Conservative} is feasible for the original problem, i.e. \eqref{Eq_Problem_R_RS}.
However, global optimality (w.r.t the conservative problem) cannot be guaranteed  due to  non-convexity.
Despite sub-optimality, such algorithm was shown to achieve good performances \cite{Tajer2011,Joudeh2016}.
\begin{algorithm}%
[t]
\caption{Conservative WMMSE Algorithm}
\label{Algthm_AO_Conservative}
\begin{algorithmic}[1]
\State Input: $P_{\mathrm{t}}$
\State Initialize: $n\gets 0$, $\widehat{R}_{\mathrm{t}}^{(n)} \gets 0$, $\mathbf{P}$
\Repeat
    \State $n\gets n+1$
    \State $\big(\widehat{g}_{\mathrm{c},k},\widehat{g}_{k}\big)
    \gets \big(\arg{\widehat{\varepsilon}_{\mathrm{c},k}^{\mathrm{MMSE}}},\arg{\widehat{\varepsilon}_{k}^{\mathrm{MMSE}}}\big)$,
    $\forall k \in \mathcal{K}$
    \label{Algthm_AO_Cons_step_g}
    \State $\big(\widehat{u}_{\mathrm{c},k},\widehat{u}_{k}\big)
    \gets \big(1/\widehat{\varepsilon}_{\mathrm{c},k}^{\mathrm{MMSE}},1/\widehat{\varepsilon}_{k}^{\mathrm{MMSE}}\big)$,
    $\forall k \in \mathcal{K}$
    \label{Algthm_AO_Cons_step_u}
    \State $(\widehat{R}_{\mathrm{t}}^{(n)},\widehat{\mathbf{c}},\mathbf{P}) \gets  \arg{\widehat{\mathcal{R}}_{\mathrm{RS}}^{\mathrm{MMSE}}(P_{\mathrm{t}},\widehat{\mathbf{g}},\widehat{\mathbf{u}})}$
\Until{$\big|\widehat{R}_{\mathrm{t}}^{(n)} - \widehat{R}_{\mathrm{t}}^{(n-1)} \big| < \epsilon_{R}$}
\State Output: $\widehat{R}_{\mathrm{t}}$, $\widehat{\mathbf{c}}$ and  $\mathbf{P}$
\end{algorithmic}
\end{algorithm}
\vspace{-4.0mm}
\subsection{Conservative Approach Limitations}
\label{Subsection_Conservative_Limitations}
In our preliminary work \cite{Joudeh2016}, we observed that the conservative max-min rate performance obtained from Algorithm \ref{Algthm_AO_Conservative} may not coincide with the optimum DoF.
In particular, the RS scheme exhibits a saturating rate behaviour for non-scaling CSIT qualities.
Similar behavior was reported in \cite{Tajer2011} in the context of multi-cell transmission where the rate saturates, contradicting the DoF result. The authors of \cite{Tajer2011} regarded the saturation to the suboptimality of the AO technique.
We show that this behaviour is due to self-interference introduced by the approximation in \eqref{Eq_wc_Rates_WMSEs_LB}, preceding the AO procedure.

It is well understood that the worst-case approach adopted in this paper does not give any consideration to the statistical distribution of the CSIT errors.
To facilitate the calculation of upper-bounds on the conservative worst-case rates in \eqref{Eq_wc_Rates_WMSEs_LB}, let us assume that $\widetilde{\mathbf{h}}_{k}$ is drawn from an arbitrary zero-mean isotropic distribution defined over an origin-centered ball with radius $\delta_{k}$.
Hence,
$\mathrm{E}\big\{\widetilde{\mathbf{h}}_{k}\big\} = \mathbf{0}$ and
$\mathrm{E}\big\{\widetilde{\mathbf{h}}_{k} \widetilde{\mathbf{h}}_{k}^{H} \big\} = \sigma_{\mathrm{e},k}^{2}\mathbf{I}$.
For CSIT that scales with SNR, $\sigma_{\mathrm{e},k}^{2}$  scales as $O(P_{\mathrm{t}}^{-\alpha_{k}})$, which follows from $\sigma_{\mathrm{e},k}^{2} = \mathrm{E}\big\{ \| \widetilde{\mathbf{h}}_{k}\|^{2} \big\}/N_{\mathrm{t}}$ and
$\| \widetilde{\mathbf{h}}_{k}\|^{2} = O(\delta_{k}^{2})$.
Note that the isotropic assumption is made for the sake of the following worst-case analysis, and does not restrict the actual CSIT errors to such distributions, as they may assume non-isotropic distributions confined within the uncertainty region.
For a given channel estimate $\widehat{\mathbf{h}}_{k}$, averaging the receive power in \eqref{Eq_T_c_k} over the introduced distribution of $\widetilde{\mathbf{h}}_{k}$ yields
\begin{equation}
\label{Eq_T_c_k_hat}
\widehat{T}_{\mathrm{c},k} = \mathbf{p}_{\mathrm{c}}^{H}\mathbf{R}_{k}\mathbf{p}_{\mathrm{c}} +
\underbrace{ \mathbf{p}_{k}^{H}\mathbf{R}_{k}\mathbf{p}_{k} + \overbrace{\sum_{i\neq k} \mathbf{p}_{i}^{H}\mathbf{R}_{k}\mathbf{p}_{i} + \sigma_{\mathrm{n}}^{2}}^{\widehat{I}_{k}}}_{\widehat{I}_{\mathrm{c},k} = \widehat{T}_{k}}
\end{equation}
where $\mathbf{R}_{k} \triangleq \mathrm{E}_{\widetilde{\mathbf{h}}_{k} \mid \widehat{\mathbf{h}}_{k} }\big\{ \mathbf{h}_{k} \mathbf{h}_{k}^{H} \big\} = \widehat{\mathbf{h}}_{k}\widehat{\mathbf{h}}_{k}^{H} + \sigma_{\mathrm{e},k}^{2} \mathbf{I}$.
\newtheorem{Lemma_R_hat_UB}[Lemma_Counter]{Lemma}
\begin{Lemma_R_hat_UB}\label{Lemma_R_hat_UB}
\textnormal{
The conservative worst-case rates in \eqref{Eq_wc_Rates_WMSEs_LB} are upper-bounded by
\begin{equation}
\label{Eq_R_hat_UB}
\widehat{R}_{\mathrm{c},k} \leq \log_{2}\big(1+ \widehat{\gamma}_{\mathrm{c},k} \big)
\quad \text{and} \quad
\widehat{R}_{k} \leq \log_{2}\big(1+ \widehat{\gamma}_{k} \big)
\end{equation}
where $\widehat{\gamma}_{\mathrm{c},k}$ and $\widehat{\gamma}_{\mathrm{c},k}$ are upper-bounds on the equivalent conservative worst-case SINRs defined as
\begin{equation}
\label{Eq_gamma_hat}
\widehat{\gamma}_{\mathrm{c},k} \triangleq |\mathbf{p}_{\mathrm{c}}^{H}\widehat{\mathbf{h}}_{k}|^{2}
\big(\widehat{I}_{\mathrm{c},k} + \widetilde{I}_{\mathrm{c},k} \big)^{-1}
\ \text{and} \
\widehat{\gamma}_{k} \triangleq | \mathbf{p}_{k}^{H}\widehat{\mathbf{h}}_{k} |^{2}
\big(\widehat{I}_{k} + \widetilde{I}_{k}   \big)^{-1}
\end{equation}
with $\widetilde{I}_{\mathrm{c},k} =  \sigma_{\mathrm{e},k}^{2} \|\mathbf{p}_{\mathrm{c}}\|^{2}$ and
$\widetilde{I}_{k} =  \sigma_{\mathrm{e},k}^{2} \|\mathbf{p}_{k}\|^{2}$.}
\end{Lemma_R_hat_UB}
The proof is given in Appendix \ref{Appendix_Proof_Lemma_R_hat_UB}.
From \eqref{Eq_gamma_hat}, it can be seen that the useful signal power components in $\widehat{\gamma}_{\mathrm{c},k}$ and $\widehat{\gamma}_{k}$ only consist of parts incorporating the channel estimate, while the parts of the desired signal power incorporating the CSIT errors, namely $\widetilde{I}_{\mathrm{c},k}$ and $\widetilde{I}_{k}$, appear as interference.
This is explained as follows:
by removing the dependencies of equalizers (and weights) on the actual channel, the robust rates (as guaranteed by the BS) are optimized while ignoring the availability of perfect CSIR,
and hence treating parts of the desired signal which incorporate $\widetilde{\mathbf{h}}_{k}$ as noise.
This yields the self-interference terms appearing in \eqref{Eq_gamma_hat}, which scale as $\widetilde{I}_{\mathrm{c},k} = O\big(P_{\mathrm{t}}^{a_{\mathrm{c}} - \alpha_{k}}\big)$ and
$\widetilde{I}_{k} = O\big(P_{\mathrm{t}}^{a_{k} - \alpha_{k}}\big)$.
This is detrimental to the achievable worst-case rates as perceived by the BS. For example, fixed CSIT qualities give $\widetilde{I}_{\mathrm{c},k}$ and $\widetilde{I}_{k}$ that scale as $O\big(P_{\mathrm{t}}^{a_{\mathrm{c}}}\big)$ and $O\big(P_{\mathrm{t}}^{a_{k}}\big)$ respectively, resulting in saturating SINRs in \eqref{Eq_gamma_hat} as $P_{\mathrm{t}}\rightarrow \infty$, which explains the behavior in \cite{Joudeh2016,Tajer2011}.
\section{Cutting-Set Method}
\label{Section_Cutting_Set}
%
In this section, we propose an algorithm that avoids conservative approximations by directly optimizing the worst-case achievable rates.
This algorithm employs an iterative procedure, known as the cutting-set method \cite{Mutapcic2009}, which switches between two steps in each iteration: optimization and pessimization (worst-case analysis).
In the optimization step, a sampled version of the semi-infinite problem is solved over finite subsets of the uncertainty regions.
In the pessimization step, worst-case analysis is carried out where channels that violate the constraints are determined and appended to the uncertainty subsets.
For the $i$th iteration, let $\overline{\mathbb{H}}_{k}\triangleq \left\{ \mathbf{h}_{k}^{(1)},\ldots, \mathbf{h}_{k}^{(i_{k})} \right\}$ and $\overline{\mathbb{H}}_{\mathrm{c},k}\triangleq \left\{ \mathbf{h}_{\mathrm{c},k}^{(1)},\ldots, \mathbf{h}_{\mathrm{c},k}^{(i_{\mathrm{c},k})} \right\}$
be the $k$th user's discretized channel uncertainty regions for the private rate and the common rate respectively.
We define the index sets
$\overline{\mathcal{I}}_{k}^{(i)}  \triangleq \{1,\ldots,i_{k}\}$ and $\overline{\mathcal{I}}_{\mathrm{c},k}^{(i)} \triangleq \{1,\ldots,i_{\mathrm{c},k}\}$, where $|\overline{\mathcal{I}}_{k}^{(i)} |,|\overline{\mathcal{I}}_{\mathrm{c},k}^{(i)}| \leq i$ as a maximum of one channel vector per set is added in each iteration.
Although $R_{k}$ and $R_{\mathrm{c},k}$ are functions of the same channel vector $(\mathbf{h}_{k})$, the corresponding uncertainty region is sampled twice where the worst-case analysis is carried out individually.
This is due to the fact that the private and common messages are independently decoded, and hence worst-case constraints should be satisfied for each case.
\subsection{Cutting-Set Algorithm}
For the $i$th iteration, the sampled problem writes as
\begin{equation}
\label{Eq_Problem_R_RS_Samp}
\mathcal{R}_{\mathrm{RS}}^{(i)}(P_{\mathrm{t}}) \! : \!
\begin{cases}
       \underset{ \bar{R}_{\mathrm{t}},\bar{\mathbf{c}},\mathbf{P} }{\max} \bar{R}_{\mathrm{t}} \\
       \text{s.t.} \ R_{k}\big(\mathbf{h}_{k}^{(j)}\big) \! + \! \bar{C}_{k}  \geq  \bar{R}_{\mathrm{t}}, \forall j \in \overline{\mathcal{I}}_{k}^{(i)},  k \in  \mathcal{K} \\
             R_{\mathrm{c},k}\big(\mathbf{h}_{\mathrm{c},k}^{(j_{\mathrm{c}})}\big) \geq \sum_{l=1}^{K} \bar{C}_{l},
            \forall j_{\mathrm{c}} \in \overline{\mathcal{I}}_{\mathrm{c},k}^{(i)},  k\in\mathcal{K} \\
            \bar{C}_{k} \geq 0, \; \forall k \in \mathcal{K}\\
            \mathrm{tr}\big(\mathbf{P}\mathbf{P}^{H}\big) \leq P_{\mathrm{t}}.
\end{cases}
\end{equation}
The optimization step involves solving \eqref{Eq_Problem_R_RS_Samp} yielding the point
$\big( \bar{R}_{\mathrm{t}}^{(i)}, \bar{\mathbf{c}}^{(i)}, \mathbf{P}^{(i)} \big)$.
Pessimization is then carried out to determine the channel vectors under which the rate constraints are most violated.
The worst-case channels corresponding to the $k$th user's rates are obtained as
\begin{equation}
\label{Eq_worst_case_channels}
\mathbf{h}_{k}^{\ast} \! = \! \arg \! \min_{\mathbf{h}_{k} \in \mathbb{H}_{k}}
\! \! \!
R_{k}^{(i)}(\mathbf{h}_{k})
\; \text{and} \;
\mathbf{h}_{\mathrm{c},k}^{\ast} \! = \! \arg \! \min_{\mathbf{h}_{\mathrm{c},k} \in \mathbb{H}_{k}}
\! \! \!
R_{\mathrm{c},k}^{(i)}(\mathbf{h}_{\mathrm{c},k})
\end{equation}
where $R_{k}^{(i)}$ and $R_{\mathrm{c},k}^{(i)}$ denote the rates when $\mathbf{P}^{(i)}$ is employed.
The rate constraints in \eqref{Eq_Problem_R_RS_Samp} are examined under the worst-case channels from \eqref{Eq_worst_case_channels}.
If $ R_{k}^{(i)}\big(\mathbf{h}_{k}^{\ast}\big) + \bar{C}_{k}^{(i)}  <  \bar{R}_{\mathrm{t}}^{(i)}$, $\mathbf{h}_{k}^{\ast}$ is appended to
$\overline{\mathbb{H}}_{k}$.
In a similar manner, if $ R_{\mathrm{c},k}^{(i)}\big(\mathbf{h}_{\mathrm{c},k}^{\ast}\big) <  \sum_{l=1}^{K} \bar{C}_{l}^{(i)}$,
$\mathbf{h}_{\mathrm{c},k}^{\ast}$ is appended to $\overline{\mathbb{H}}_{\mathrm{c},k}$.
The cutting-set procedure is summarized in Algorithm \ref{Algthm_Cutting_Set}.
Defining the rate violations after the $i$th pessimization as
$V^{(i)}_{k} \triangleq \bar{R}_{\mathrm{t}}^{(i)} - R_{k}^{(i)}\big(\mathbf{h}_{k}^{\ast}\big) - \bar{C}_{k}^{(i)}$ and
$V^{(i)}_{\mathrm{c},k} \triangleq \sum_{l\in\mathcal{K}}\bar{C}_{l}^{(i)} - R_{\mathrm{c},k}^{(i)}\big(\mathbf{h}_{\mathrm{c},k}^{\ast}\big)$,
the stopping criteria in Algorithm \ref{Algthm_Cutting_Set} is specified as a maximum violation,
i.e.
\begin{equation}
\nonumber
\max_{k} \left\{ \max \left\{ V^{(i)}_{k}, V^{(i)}_{\mathrm{c},k} \right\} \right\}_{k \in \mathcal{K}} \leq \epsilon_{V}
\end{equation}
where $\epsilon_{V} > 0$ is some arbitrary tolerance constant.

The cutting-set algorithm converges to the optimum solution of the original problem given that the optimization and pessimization steps are solved to global optimality in each iteration \cite{Mutapcic2009}.
However, the optimization step here involves solving the non-convex problem in \eqref{Eq_Problem_R_RS_Samp}, and hence global optimality may not be guaranteed in general.
Alternatively, a stationary solution can be guaranteed as follows.
\newtheorem{Proposition_Cutting_Set_Conv}[Proposition_Counter]{Proposition}
\begin{Proposition_Cutting_Set_Conv}\label{Proposition_Cutting_Set_Conv}
\textnormal{Given that the optimization step in Algorithm \ref{Algthm_Cutting_Set} yields a KKT point of the corresponding sampled problem in \eqref{Eq_Problem_R_RS_Samp}, and the pessimization step is exact (i.e. the global minimizers of \eqref{Eq_worst_case_channels} are obtained), then the iterates generated by  Algorithm \ref{Algthm_Cutting_Set} converge to the set of KKT points of the semi-infinite problem in \eqref{Eq_Problem_R_RS}.
}
\end{Proposition_Cutting_Set_Conv}
The proof of Proposition \ref{Proposition_Cutting_Set_Conv} employs ideas from \cite{Wu2005}, where iterative methods (that coincide with the cutting-set algorithm) are proposed to solve non-linear semi-infinite programs. Details of the proof are relegated to Appendix \ref{Appendix_Proof_Proposition_Cutting_Set_Conv}.
In the following, the optimization and pessimization steps are thoroughly addressed.
For ease of notation, the superscript $(i)$ is dropped from the variables where is its understood that optimization and pessimization are part of a given iteration.
\begin{algorithm}[t]
\caption{Cutting-Set method.}
\label{Algthm_Cutting_Set}
\begin{algorithmic}[1]
\State Input: $P_{\mathrm{t}}$
\State Initialize: $i \! \gets \! 0$, $i_{k},i_{\mathrm{c},k} \! \gets \! 1$ and
$\overline{\mathbb{H}}_{k},\overline{\mathbb{H}}_{\mathrm{c},k} \! \gets \! \big\{ \! \widehat{\mathbf{h}}_{k}  \! \big\},\forall k \! \in \! \mathcal{K}$
\Repeat
    \State $i \gets i+1$
    \Statex Optimization:
    \State $\big(\bar{R}_{\mathrm{t}}^{(i)},\bar{\mathbf{c}}^{(i)},\mathbf{P}^{(i)}\big) \gets
    \arg{\mathcal{R}^{(i)}(P_{\mathrm{t}})}$
    \Statex Pessimization:
    \State For all $k\in \mathcal{K}$, \textbf{do}
    \State Obtain $\mathbf{h}_{k}^{\ast},\mathbf{h}_{\mathrm{c},k}^{\ast}$ by solving \eqref{Eq_worst_case_channels}
    \If {$ R_{k}^{(i)}\big(\mathbf{h}_{k}^{\ast}\big) + \bar{C}_{k}^{(i)} \! < \!\bar{R}_{\mathrm{t}}^{(i)}$}
    \State $i_{k} \gets i_{k}+1$ and $\mathbf{h}_{k}^{(i_{k})} \gets \mathbf{h}_{k}^{\ast}$
    \State  $\overline{\mathbb{H}}_{k} \gets \big\{ \overline{\mathbb{H}}_{k},\mathbf{h}_{k}^{(i_{k})} \big\}$
    \EndIf
    \If {$R_{\mathrm{c},k}^{(i)}\big(\mathbf{h}_{k}^{\ast}\big) <  \sum_{l=1}^{K}\bar{C}_{l}^{(i)}$}
    \State $i_{\mathrm{c},k}  \gets  i_{\mathrm{c},k}  + 1$ and
    $\mathbf{h}_{\mathrm{c},k}^{(i_{\mathrm{c},k})}  \gets  \mathbf{h}_{\mathrm{c},k}^{\ast}$
    \State $\overline{\mathbb{H}}_{\mathrm{c},k}  \gets  \big\{\overline{\mathbb{H}}_{\mathrm{c},k},\mathbf{h}_{\mathrm{c},k}^{(i_{\mathrm{c},k})}  \big\}$
    \EndIf
\Until{stopping criteria is met}
\label{Algthm_Cutting_Set_stopping_criteria}
\State Output: $\mathbf{P}^{(i)}$, $\bar{R}_{\mathrm{t}}^{(i)}$ and $\bar{\mathbf{c}}^{(i)}$
\end{algorithmic}
\end{algorithm}
%
\subsection{Optimization}
\label{subsection_optimization}
The Rate-WMMSE relationship in Section \ref{Subsection_Rate_WMSE} is revisited to transform the sampled problem \eqref{Eq_Problem_R_RS_Samp} into an equivalent WMSE problem formulated as
\begin{align}
\label{Eq_Problem_WMSE_R_RS_Samp}
&\bar{\mathcal{R}}_{\mathrm{RS}}^{(i)}(P_{\mathrm{t}}):  \\
\nonumber
&
\begin{cases}
       \underset{\bar{R}_{\mathrm{t}},\bar{\mathbf{c}},\mathbf{P},\mathbf{g},\mathbf{u}}{\max}  \bar{R}_{\mathrm{t}} \\
       \text{s.t.} \\
       1 \! - \! \xi_{k}\big(\mathbf{h}_{k}^{(j)},g_{k}^{(j)},u_{k}^{(j)}\big) \! + \! \bar{C}_{k}
         \! \geq \! \bar{R}_{\mathrm{t}}, \forall j \in \overline{\mathcal{I}}_{k}^{(i)}, k \in  \mathcal{K} \\
      1 \! - \! \xi_{\mathrm{c},k}\big(\mathbf{h}_{\mathrm{c},k}^{(j_{\mathrm{c}})},
       g_{\mathrm{c},k}^{(j_{\mathrm{c}})},u_{\mathrm{c},k}^{(j_{\mathrm{c}})}\big) \geq \sum_{l=1}^{K} \bar{C}_{l},
       \forall j_{\mathrm{c}} \in \overline{\mathcal{I}}_{\mathrm{c},k}^{(i)}, k \in  \mathcal{K} \\
       \bar{C}_{k} \geq 0, \; \forall k \in \mathcal{K}\\
       \mathrm{tr}\big(\mathbf{P}\mathbf{P}^{H}\big) \leq P_{\mathrm{t}}.
\end{cases}
\end{align}
$\mathbf{g} \triangleq \{\mathbf{g}_{\mathrm{c},k},\mathbf{g}_{k} \mid k \in \mathcal{K} \}$ is the sampled set of equalizers with $\mathbf{g}_{\mathrm{c},k} \triangleq \left\{ g_{\mathrm{c},k}^{(j_{\mathrm{c}} )} \mid  j_{\mathrm{c}} \in \overline{\mathcal{I}}_{\mathrm{c},k}^{(i)} \right\}$ and
$\mathbf{g}_{k} \triangleq \left\{ g_{k}^{(j)} \mid j \in \overline{\mathcal{I}}_{k}^{(i)} \right\}$,
while
$\mathbf{u} \triangleq \{\mathbf{u}_{\mathrm{c},k},\mathbf{u}_{k} \mid k \in \mathcal{K} \}$
is the sampled set of weights where
$\mathbf{u}_{\mathrm{c},k} \triangleq \left\{ u_{\mathrm{c},k}^{(j_{\mathrm{c}})} \mid  j_{\mathrm{c}} \in \overline{\mathcal{I}}_{\mathrm{c},k}^{(i)} \right\}$
and
$\mathbf{u}_{k} \triangleq \left\{ u_{k}^{(j)} \mid j \in \overline{\mathcal{I}}_{k}^{(i)} \right\}$.
Contrary to the conservative approach, the sampling of the equalizers and weights captures their dependencies on the actual channel, which reflects the availability of perfect CSIR.

The AO principle used in Section \ref{Subsection_AO_conservative} is employed to solve the sampled problem.
For a given iteration, $\mathbf{g}$ is firstly optimized by applying the MMSE solution in Section \ref{Subsection_Rate_WMSE} to each equalizer in the sampled set, i.e.
$g_{\mathrm{c},k}^{(j_{\mathrm{c}})}=g_{\mathrm{c},k}^{\mathrm{MMSE}(j_{\mathrm{c}})}$ and
$g_{k}^{(j)}=g_{k}^{\mathrm{MMSE}(j)}$ for all $j_{\mathrm{c}}$, $j$ and $k$.
The optimality of this step comes from the optimality of the MMSE solution for each
$\xi_{\mathrm{c},k}\big(\mathbf{h}_{\mathrm{c},k}^{(j_{\mathrm{c}})},g_{\mathrm{c},k}^{(j_{\mathrm{c}})}\big)$
and
$\xi_{k}\big(\mathbf{h}_{k}^{(j)},g_{k}^{(j)}\big)$.
Next, $\mathbf{u}$ is optimized in a similar manner using the solution in Section \ref{Subsection_Rate_WMSE},  i.e.
$u_{\mathrm{c},k}^{(j_{\mathrm{c}})}=u_{\mathrm{c},k}^{\mathrm{MMSE}(j_{\mathrm{c}})}$ and
$u_{k}^{(j)}=u_{k}^{\mathrm{MMSE}(j)}$.
Finally, $(\bar{R}_{\mathrm{t}},\bar{\mathbf{c}},\mathbf{P})$ are updated by solving
problem \eqref{Eq_Problem_WMSE_R_RS_Samp} for fixed $\mathbf{g}$ and $\mathbf{u}$.
The resulting problem is convex with a linear objective function and quadratic and linear constraints, and can be efficiently solved using interior-point methods \cite{Boyd2004}.
Following the same argument in Section \ref{Subsection_AO_conservative}, the AO algorithm described here is guaranteed to converge.
Furthermore, the KKT optimality of the generated solution is established in the following result.
%
\newtheorem{Proposition_AO_Stationary}[Proposition_Counter]{Proposition}
\begin{Proposition_AO_Stationary}\label{Proposition_AO_Stationary}
\textnormal{The iterates generated by the AO procedure described above converge to the set of KKT points of the $i$th sampled rate problem in \eqref{Eq_Problem_R_RS_Samp}.
}
\end{Proposition_AO_Stationary}
The convergence of WMMSE algorithms to stationary (KKT) points was established for the sum-rate problem \cite{Shi2011} and the max-min fair problem \cite{Razaviyayn2013b} in the context of the MIMO Interfering BC (IBC) under prefect CSI and NoRS.
It was later shown that the WMMSE algorithm belongs to a class of inexact BCDs, known as Successive Upper-bound Minimization (SUM), and more general analysis and proofs were established in \cite{Razaviyayn2013,Razaviyayn2014}.
The proof of Proposition \ref{Proposition_AO_Stationary} is based on \cite{Razaviyayn2014}, and is relegated to Appendix \ref{Appendix_Proof_Proposition_AO_Stationary}.
Due to non-convexity, the KKT point obtained by the AO algorithm may be suboptimal.
However, the effectiveness of this algorithm is demonstrated through simulation results.
\subsection{Pessimization}
For the outputs of the optimization step, the pessimization step determines whether the rate constraints are violated under an updated set of worst-case channels.
This involves solving the problems in \eqref{Eq_worst_case_channels}, which can be formulated in terms of the MMSEs due to their monotonic relationship with the rates.
The worst-case MMSEs are defined as
\begin{subequations}
\label{Eq_worst_case_MMSE}
\begin{align}
\label{Eq_worst_case_MMSE_p}
\bar{\varepsilon}_{k}^{\mathrm{MMSE}} & \triangleq
\max_{\mathbf{h}_{k} \in \mathbb{H}_{k}} \varepsilon_{k}^{\mathrm{MMSE}}(\mathbf{h}_{k})
\\
\label{Eq_worst_case_MMSE_c}
\bar{\varepsilon}_{\mathrm{c},k}^{\mathrm{MMSE}} & \triangleq
\max_{\mathbf{h}_{\mathrm{c},k} \in \mathbb{H}_{k}} \varepsilon_{\mathrm{c},k}^{\mathrm{MMSE}}(\mathbf{h}_{\mathrm{c},k}).
\end{align}
\end{subequations}
The private rate constraint is violated if we have
$\bar{\varepsilon}_{k}^{\mathrm{MMSE}} > 2^{-(\bar{R}_{\mathrm{t}} - \bar{C}_{k})}$,
and the common rate constraint is violated given that
$\bar{\varepsilon}_{\mathrm{c},k}^{\mathrm{MMSE}}  > 2^{-\bar{R}_{\mathrm{c}}}$.
Solving the problems \eqref{Eq_worst_case_MMSE} involves maximizing non-linear fractional functions over compact convex sets.
Such problems can be solved using Dinkelbach's method \cite{Dinkelbach1967}, where the fractional program is transformed into a parametric auxiliary problem solved iteratively.
Next, the employment of this method is further explained.
\newtheorem{Lemma_Dinkelbach}[Lemma_Counter]{Lemma}
\begin{Lemma_Dinkelbach}\label{Lemma_Dinkelbach}
\textnormal{
$\mathbf{h}_{k}^{\ast}$ and $\mathbf{h}_{\mathrm{c},k}^{\ast}$ are the global maximizers of problems \eqref{Eq_worst_case_MMSE_p} and \eqref{Eq_worst_case_MMSE_c} respectively if and only if they are the global maximizers of the parametric problems formulated as
\begin{equation}
\nonumber
\mathcal{D}_{k}( \! \lambda_{k} \! ) \! \! : \! \! \max_{\mathbf{h}_{k} \in \mathbb{H}_{k}} \! \! I_{k} - \lambda_{k}T_{k}
\ \ \text{and} \ \
\mathcal{D}_{\mathrm{c},k}( \! \lambda_{\mathrm{c},k} \! ) \! \! : \!  \! \max_{\mathbf{h}_{\mathrm{c},k} \in \mathbb{H}_{k}} \! \! \!
I_{\mathrm{c},k} - \lambda_{\mathrm{c},k}T_{\mathrm{c},k}
\end{equation}
respectively, when the parameters $\lambda_{k} $ and $\lambda_{\mathrm{c},k}$ are given by
$\lambda_{k}^{\ast} \triangleq \bar{\varepsilon}_{k}^{\mathrm{MMSE}}$ and
$\lambda_{\mathrm{c},k}^{\ast} \triangleq \bar{\varepsilon}_{\mathrm{c},k}^{\mathrm{MMSE}}$ respectively.}
\end{Lemma_Dinkelbach}
Lemma \ref{Lemma_Dinkelbach} follows directly from the  theorem in \cite{Dinkelbach1967}.
It can be seen that $\mathcal{D}_{k}(\lambda_{k}^{\ast}) , \mathcal{D}_{\mathrm{c},k}(\lambda_{\mathrm{c},k}^{\ast}) = 0$.
Hence, solving the pessimization problems in  \eqref{Eq_worst_case_MMSE} is equivalent to finding the zeros of their corresponding parametric auxiliary problems in Lemma \ref{Lemma_Dinkelbach}.
This can be achieved using Dinkelbach's iterative algorithm \cite{Dinkelbach1967}.
Note that it is commonly assumed that the fractional program is concave-convex, i.e. with a concave numerator and a convex denominator, yielding
convex auxiliary problems.
Nevertheless, it follows from \cite{Dinkelbach1967} that this assumption is not necessary as long as the auxiliary problem can be solved
to global optimality for a given parameter.
The auxiliary problems in Lemma \ref{Lemma_Dinkelbach} are rewritten as
\begin{align}
\nonumber
& \mathcal{D}_{k}(\lambda_{k}): \\
\label{Eq_worst_case_MMSE_k_Par}
& \max_{\mathbf{h}_{k} \in \mathbb{H}_{k}}
\mathbf{h}_{k}^{H}
\underbrace{\big( (1-\lambda_{k})\bar{\mathbf{Q}}_{k} - \lambda_{k}\mathbf{Q}_{k} \big)}_{\mathbf{A}_{k}(\lambda_{k})}
\mathbf{h}_{k} + (1-\lambda_{k})\sigma_{\mathrm{n}}^{2} \\
\nonumber
& \mathcal{D}_{\mathrm{c},k}(\lambda_{\mathrm{c},k}): \\
\label{Eq_worst_case_MMSE_c_Par}
& \max_{\mathbf{h}_{\mathrm{c},k} \in \mathbb{H}_{k}}
\! \!
\mathbf{h}_{\mathrm{c},k}^{H} \!
\underbrace{\big( (1-\lambda_{\mathrm{c},k})\mathbf{Q}_{\mathrm{p}} \! - \! \lambda_{\mathrm{c},k}\mathbf{Q}_{\mathrm{c}} \big)}_{\mathbf{A}_{\mathrm{c},k}(\lambda_{\mathrm{c},k})}
\!
\mathbf{h}_{\mathrm{c},k} \! + \! (1 \! - \! \lambda_{\mathrm{c},k})\sigma_{\mathrm{n}}^{2}
\end{align}
where $\mathbf{Q}_{k} \triangleq \mathbf{p}_{k}\mathbf{p}_{k}^{H}$,
$\mathbf{Q}_{\mathrm{c}} \triangleq \mathbf{p}_{\mathrm{c}}\mathbf{p}_{\mathrm{c}}^{H}$,
$\mathbf{Q}_{\mathrm{p}} \triangleq \sum_{k=1}^{K}\mathbf{Q}_{k}$, and
$\bar{\mathbf{Q}}_{k} \triangleq \mathbf{Q}_{\mathrm{p}} - \mathbf{Q}_{k}$.
This follows from substituting the receive power and interference expressions in \eqref{Eq_T_c_k}.
For given parameters, \eqref{Eq_worst_case_MMSE_k_Par} and \eqref{Eq_worst_case_MMSE_c_Par} are Quadratically Constrained Quadratic Programs (QCQPs),
where
$\mathbf{A}_{k}(\lambda_{k})$ and $\mathbf{A}_{\mathrm{c},k}(\lambda_{\mathrm{c},k})$ are symmetric and possibly
indefinite\footnote{Updating the parameters using Dinkelbach's algorithm, we have $\lambda_{k},\lambda_{\mathrm{c},k} \in  [0,1]$. $\mathbf{A}_{k}(0),\mathbf{A}_{\mathrm{c},k}(0) \succeq 0$, while $\mathbf{A}_{k}(1), \mathbf{A}_{\mathrm{c},k}(1) \preceq 0$. Otherwise, they are generally indefinite.}.
Hence, \eqref{Eq_worst_case_MMSE_k_Par} and \eqref{Eq_worst_case_MMSE_c_Par} are non-convex optimization problems in general.
For this reason, we resort to relaxation. First, we introduce the matrix variables
$\mathbf{X}_{k} =\mathbf{h}_{k}\mathbf{h}_{k}^{H}$ and $\mathbf{X}_{\mathrm{c},k} = \mathbf{h}_{\mathrm{c},k}\mathbf{h}_{\mathrm{c},k}^{H}$
from which the quadratic terms in \eqref{Eq_worst_case_MMSE_k_Par} and \eqref{Eq_worst_case_MMSE_c_Par} are eliminated by writing
$\mathbf{h}_{k}^{H}\mathbf{A}_{k}\mathbf{h}_{k} = \mathrm{tr} \big(\mathbf{X}_{k} \mathbf{A}_{k}\big)$
and
$\mathbf{h}_{\mathrm{c},k}^{H}\mathbf{A}_{\mathrm{c},k}\mathbf{h}_{\mathrm{c},k} = \mathrm{tr} \big(\mathbf{X}_{\mathrm{c},k} \mathbf{A}_{\mathrm{c},k}\big)$.
Next, the equalities associated with the introduced matrices are relaxed into inequalities such that $\mathbf{X}_{k} \succeq \mathbf{h}_{k}\mathbf{h}_{k}^{H}$ and $\mathbf{X}_{\mathrm{c},k} \succeq \mathbf{h}_{\mathrm{c},k}\mathbf{h}_{\mathrm{c},k}^{H}$.
The resulting relaxed problems are formulated as
\begin{equation}
\label{Eq_worst_case_MMSE_k_Par_SDR}
\mathcal{D}_{k}^{\mathrm{r}}(\lambda_{k}):
\begin{cases}
       \underset{\mathbf{X}_{k},\mathbf{h}_{k}}{\max} & \mathrm{tr}\big(\mathbf{X}_{k}\mathbf{A}_{k}(\lambda_{k})\big) + (1-\lambda_{k})\sigma_{\mathrm{n}}^{2} \\
       \text{s.t.}
                    & \! \! \! \mathrm{tr}(\mathbf{X}_{k}) - 2\Re(\mathbf{h}_{k}^{H} \widehat{\mathbf{h}}_{k}) + \widehat{\mathbf{h}}_{k}^{H}\widehat{\mathbf{h}}_{k} \leq \delta_{k}^{2} \\
                    & \left[
                        \begin{array}{cc}
                          \mathbf{X}_{k} & \mathbf{h}_{k} \\
                          \mathbf{h}_{k}^{H} & 1 \\
                        \end{array}
                      \right] \succeq  0
\end{cases}
\end{equation}
\begin{equation}
\label{Eq_worst_case_MMSE_c_Par_SDR}
\! \!   \mathcal{D}_{\mathrm{c},k}^{\mathrm{r}}(\! \lambda_{\mathrm{c},k} \!) \! : \! \!
\begin{cases}
        \! \underset{\mathbf{X}_{\mathrm{c},k},\mathbf{h}_{\mathrm{c},k}}{\max} \!  \mathrm{tr}\big(\mathbf{X}_{\mathrm{c},k}\mathbf{A}_{\mathrm{c},k}(\lambda_{\mathrm{c},k})\big) \! + \! ( \! 1 \!- \!\lambda_{\mathrm{c},k} \! )\sigma_{\mathrm{n}}^{2} \\
       \text{s.t.} \
       \mathrm{tr}(\mathbf{X}_{\mathrm{c},k}) \! - \!  2\Re(\mathbf{h}_{\mathrm{c},k}^{H} \widehat{\mathbf{h}}_{k})  \! + \! \widehat{\mathbf{h}}_{k}^{H}\widehat{\mathbf{h}}_{k} \leq \delta_{k}^{2} \\
       \quad \quad \quad
        \left[
        \begin{array}{cc}
         \mathbf{X}_{\mathrm{c},k} & \mathbf{h}_{\mathrm{c},k} \\
          \mathbf{h}_{\mathrm{c},k}^{H} & 1 \\
          \end{array}
           \right] \succeq  0
\end{cases}
\end{equation}
where the relaxed inequalities are rewritten using the Schur Complement.
\eqref{Eq_worst_case_MMSE_k_Par_SDR} and \eqref{Eq_worst_case_MMSE_c_Par_SDR} are SDPs and can be efficiently
solved.
Due to the relaxations, the feasible sets in \eqref{Eq_worst_case_MMSE_k_Par_SDR} and \eqref{Eq_worst_case_MMSE_c_Par_SDR} contain their counterparts in \eqref{Eq_worst_case_MMSE_k_Par} and \eqref{Eq_worst_case_MMSE_c_Par}. It follows that
$\mathcal{D}_{k}^{\mathrm{r}}(\lambda_{k}) \geq \mathcal{D}_{k}(\lambda_{k})$
and
$\mathcal{D}_{\mathrm{c},k}^{\mathrm{r}}(\lambda_{\mathrm{c},k}) \geq \mathcal{D}_{\mathrm{c},k}(\lambda_{\mathrm{c},k})$.
Before proceeding to the next result, we denote the optimum solutions of \eqref{Eq_worst_case_MMSE_k_Par_SDR} and \eqref{Eq_worst_case_MMSE_c_Par_SDR} as
$\big(\mathbf{X}_{k}^{\circ},\mathbf{h}_{k}^{\circ}\big)$
and $\big(\mathbf{X}_{\mathrm{c},k}^{\circ},\mathbf{h}_{\mathrm{c},k}^{\circ}\big)$ respectively.
\newtheorem{Lemma_Tight_Relaxation}[Lemma_Counter]{Lemma}
\begin{Lemma_Tight_Relaxation}\label{Lemma_Tight_Relaxation}
\textnormal{
The relaxations in \eqref{Eq_worst_case_MMSE_k_Par_SDR} and \eqref{Eq_worst_case_MMSE_c_Par_SDR} are tight at optimality, i.e.
$\mathbf{X}_{k}^{\circ} =  \mathbf{h}_{k}^{\circ}\mathbf{h}_{k}^{\circ^{H}}$
and
$\mathbf{X}_{\mathrm{c},k}^{\circ} =  \mathbf{h}_{\mathrm{c},k}^{\circ}\mathbf{h}_{\mathrm{c},k}^{\circ^{H}}$.
As a result, $\mathbf{h}_{k}^{\circ}$ and $\mathbf{h}_{\mathrm{c},k}^{\circ}$ are optimum solutions for
\eqref{Eq_worst_case_MMSE_k_Par} and \eqref{Eq_worst_case_MMSE_c_Par} respectively.
Finally, we have $\mathcal{D}_{k}^{\mathrm{r}}(\lambda_{k}) = \mathcal{D}_{k}(\lambda_{k})$ and
$\mathcal{D}_{\mathrm{c},k}^{\mathrm{r}}(\lambda_{\mathrm{c},k}) = \mathcal{D}_{\mathrm{c},k}(\lambda_{\mathrm{c},k})$.
}
\end{Lemma_Tight_Relaxation}
Lemma \ref{Lemma_Tight_Relaxation} follows directly from  \cite[Appendix B.1]{Boyd2004},
by noting that \eqref{Eq_worst_case_MMSE_k_Par} and \eqref{Eq_worst_case_MMSE_c_Par} are QCQPs, with a single constraint each, that satisfy Slater's condition\footnote{It should be noted that each of \eqref{Eq_worst_case_MMSE_k_Par} and \eqref{Eq_worst_case_MMSE_c_Par} is also known as a trust-region subproblem, and can be solved using alternative methods \cite{Tao1998}.}.

Since \eqref{Eq_worst_case_MMSE_k_Par} and \eqref{Eq_worst_case_MMSE_c_Par} are globally solved for given parameters, Dinkelbach's algorithm can be employed.
This is carried out separately for $\varepsilon_{k}^{\mathrm{MMSE}}$ and $\varepsilon_{\mathrm{c},k}^{\mathrm{MMSE}}$ for all $k \in \mathcal{K}$, as the worst-case analysis is independent in each case.
For $\varepsilon_{k}^{\mathrm{MMSE}}$, the parameter is initialized as $\lambda_{k}^{(1)} = 2^{-(\bar{R}_{\mathrm{t}} - \bar{C}_{k})}$, and
$\mathcal{D}_{k}\big(\lambda_{k}^{(1)}\big)$ is obtained by solving the relaxed problem \eqref{Eq_worst_case_MMSE_k_Par_SDR}.
$\mathcal{D}_{k}\big(\lambda_{k}^{(1)}\big) \leq 0$ implies that $\lambda_{k}^{(1)} \geq \bar{\varepsilon}_{k}^{\mathrm{MMSE}}$,
as $\mathcal{D}_{k}\big(\bar{\varepsilon}_{k}^{\mathrm{MMSE}}\big) = 0$ and $\mathcal{D}_{k}$ is strictly decreasing in its parameter \cite{Dinkelbach1967}.
In this case, the rate constraint is not violated and there is no need to proceed.
Otherwise, the worst-case channel is updated as $\mathbf{h}_{k}^{(1)} = \arg{\mathcal{D}_{k}\big(\lambda_{k}^{(1)}\big)}$,
and the parameter to be used in the next iteration is obtained as
$\lambda_{k}^{(2)} = \varepsilon_{k}^{\mathrm{MMSE}}\big(\mathbf{h}_{k}^{(1)}\big)$.
This procedure is summarized in Algorithm \ref{Algthm_Pessim_k}, where $\epsilon_{\mathcal{D}} > 0$ determines the accuracy of the solution, and $\{\}$ corresponds to an empty set.
It follows directly from Lemma \ref{Lemma_Dinkelbach} that if the rate constraints are violated,
$\lambda_{k}^{(m)}$ and $\mathbf{h}_{k}^{(m)}$ converge to $\bar{\varepsilon}_{k}^{\mathrm{MMSE}}$ and $\mathbf{h}_{k}^{\ast}$, respectively.
For $\varepsilon_{\mathrm{c},k}^{\mathrm{MMSE}}$, the parameter is initialized as
$\lambda_{\mathrm{c},k}^{(1)} = 2^{-\bar{R}_{\mathrm{c}}}$ and the same steps are followed yielding
$\mathbf{h}_{\mathrm{c},k}^{\ast}$ if the common rate is violated, and $\{\}$ otherwise\footnote{Summary in an algorithm form is omitted to avoid repetition.}.
%
\begin{algorithm}%
[t]
\caption{Pessimization through Dinkelbach's Algorithm}
\label{Algthm_Pessim_k}
\begin{algorithmic}[1]
\State Initialize: $\lambda_{k}^{(1)} \gets 2^{-(\bar{R}_{\mathrm{t}} - \bar{C}_{k})}$ and $m\gets 0$
\Repeat
    \State $m \gets m+1$
    \State obtain $\mathcal{D}_{k}\big(\lambda_{k}^{(m)}\big)$ by solving \eqref{Eq_worst_case_MMSE_k_Par_SDR}
    \State  $\mathbf{h}_{k}^{(m)} \gets \arg{\mathcal{D}_{k}\big(\lambda_{k}^{(m)}\big)}$
    \State  $\lambda_{k}^{(m+1)} \gets \varepsilon_{k}^{\mathrm{MMSE}}\big(\mathbf{h}_{k}^{(m)}\big)$ from \eqref{Eq_MMSE_k}
\Until{$\mathcal{D}_{k}\big( \lambda_{k}^{(m)}  \big) \leq \epsilon_{\mathcal{D}}$, or $m = m_{\max}$}
\If {$m > 1$}
\State $\mathbf{h}_{k}^{\ast} \gets \mathbf{h}_{k}^{(m)}$
\State Output: $\mathbf{h}_{k}^{\ast}$
\Else
\State Output: $\{\}$
\EndIf
\end{algorithmic}
\end{algorithm}
%

Since the optimization step in the previous subsection yields a KKT point for the sampled problem, and the pessimization step in this subsection is exact, it follows from Proposition \ref{Proposition_Cutting_Set_Conv} that a KKT point for problem \eqref{Eq_Problem_R_RS} is obtained by Algorithm \ref{Algthm_Cutting_Set}.
\section{Simulation Results}
\label{Section_Simulation_Results}
In this section, the performance is assessed through simulations.
All optimization problems requiring interior-point methods are solved using the CVX toolbox \cite{Grant2008}.
A three-user system with  $K,N_{t} = 3$ is considered throughout the simulations, unless stated otherwise.
The noise variance is fixed as $\sigma_{\mathrm{n}}^{2} = 1$.
A given channel matrix $\mathbf{H}$ has i.i.d. entries drawn from the distribution $\mathcal{C}\mathcal{N}\left(0,1\right)$.
The corresponding estimate is obtained as $\widehat{\mathbf{H}} = \mathbf{H} - \widetilde{\mathbf{H}}$,
where each error vector is drawn from a uniform distribution over the corresponding uncertainty region with
$\delta_{k}^{2} = \beta_{k}P_{\mathrm{t}}^{-\alpha_{k}}$, where $\beta_{k}$ is a constant.
We consider the conservative (con) and the cutting-set (cs) methods for both the NoRS and RS strategies, yielding four different designs: NoRS-con, NoRS-cs, RS-con and RS-cs.
The NoRS-con and NoRS-cs results are obtained from Algorithm \ref{Algthm_AO_Conservative} and Algorithm \ref{Algthm_Cutting_Set} respectively, by discarding the common message. It should be noted that the NoRS-con design is equivalent to the MSE-based design in \cite{Vucic2009a}.
\subsection{Max-Min Fair Rate Performance}
\begin{figure}
\vspace{-5.0mm}
  \hspace{-0.05 in}
  \subfloat[][{$\delta = 0.05$}]
  {\label{Fig_Rate_11} \includegraphics[width = 0.25\textwidth]{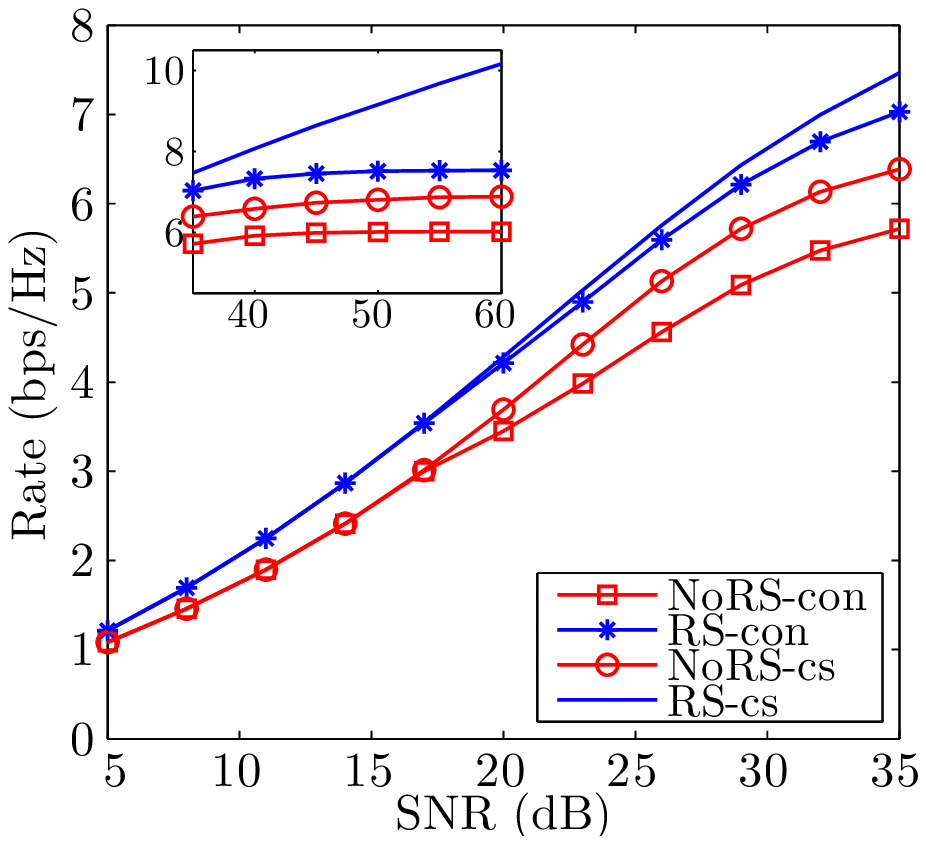}}
  \hspace{-0.2 in}
  \subfloat[][{$\delta = 0.15$}]
  {\label{Fig_Rate_12} \includegraphics[width = 0.25\textwidth]{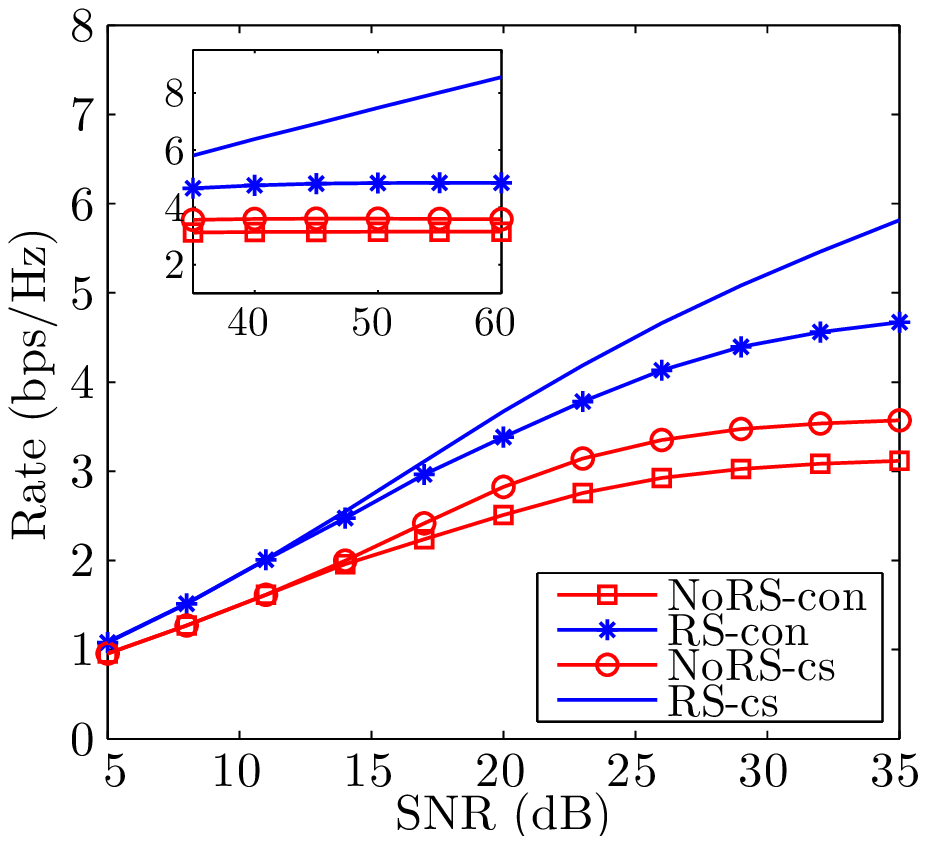}}
  \caption{Rate performance for $K,N_{\mathrm{t}} = 3$, and $\delta_{1},\delta_{2},\delta_{3} = \delta$.}
  \label{Fig_Rate_1}
\vspace{-4.0mm}
\end{figure}
\begin{figure}
  \hspace{-0.05 in}
  \subfloat[][{$\delta = 0.05$}]
  {\label{Fig_Rate_21} \includegraphics[width = 0.25\textwidth]{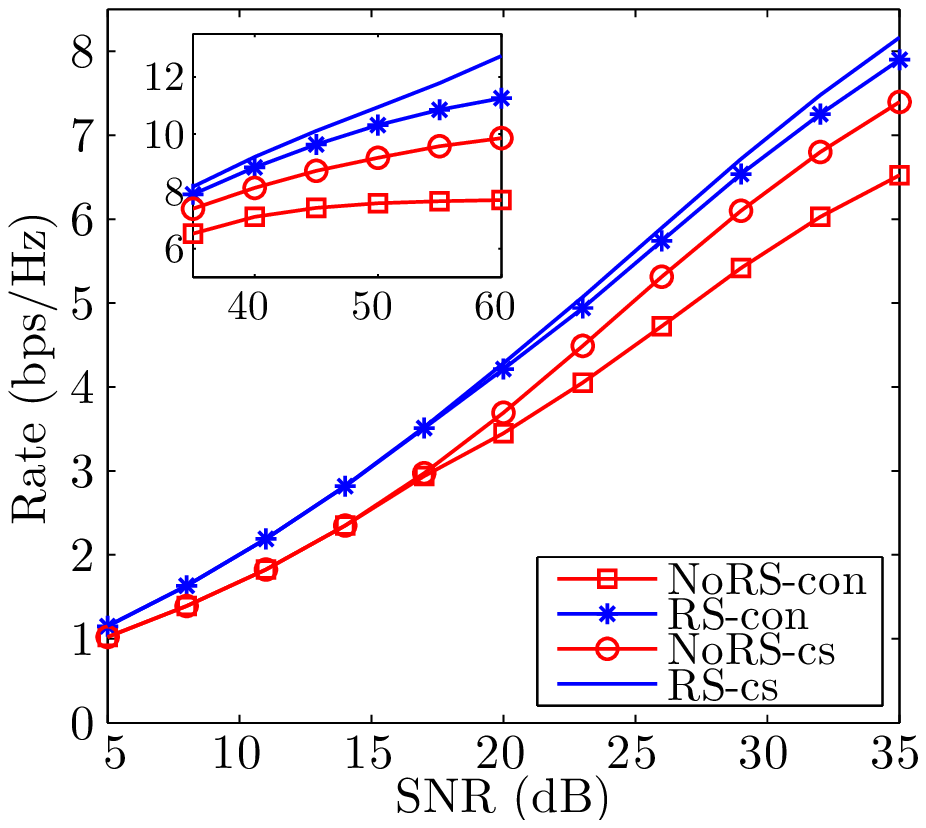}}
  \hspace{-0.2 in}
  \subfloat[][{$\delta = 0.15$}]
  {\label{Fig_Rate_22} \includegraphics[width = 0.25\textwidth]{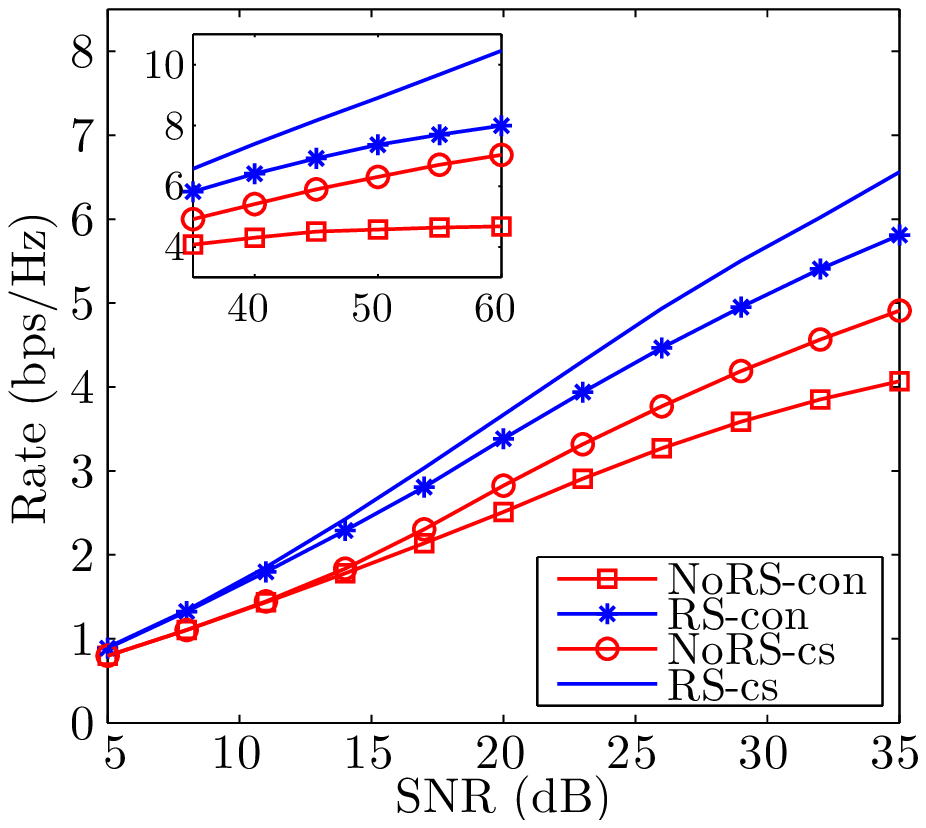}}
  \caption{Rate performance for $K ,N_{\mathrm{t}} \! = \! 3$, $\delta_1 \! = \! \delta$, and $\delta_2,\delta_3 \! = \! \delta\sqrt{10P_{\mathrm{t}}^{-0.5}}$.}
  \label{Fig_Rate_2}
\end{figure}
First, we examine the robust max-min rate performance for the four aforementioned designs.
Results for non-scaling CSIT errors (i.e. $\alpha_{1},\alpha_{2},\alpha_{3} = 0$) are shown in Fig. \ref{Fig_Rate_1} with $\delta_{1},\delta_{2},\delta_{3} = \delta$, where $\delta = 0.05$ and $0.15$ for Fig. \ref{Fig_Rate_11} and \ref{Fig_Rate_12}, respectively.
The worst-case rates are averaged over $100$ realizations of $\widehat{\mathbf{H}}$, where each estimate is obtained from an independent realization of $\mathbf{H}$.
For a given strategy (NoRS or RS), the cs design outperforms the con design, specifically in the intermediate and high SNR regimes.
This gap grows larger with increased SNR and CSIT uncertainty,
due to the increased influence of self-interference resulting from the conservative approximation.
As expected from Theorem \ref{Theorem_RS_NoRS_Max_Min_DoF}, NoRS schemes saturate as SNR grows large
($\bar{d}^{\ast} = 0$).
This trend is also followed by the RS-con design, which at first glance seems to contradict the result in Theorem \ref{Theorem_RS_NoRS_Max_Min_DoF},
yet can be explained in the light of the analysis in Section \ref{Subsection_Conservative_Limitations}.
On the other hand, the RS-cs design coincides with the result in \eqref{Eq_Max_Min_DoF_RS} and achieves an ever growing rate performance with an approximate DoF\footnote{Obtained by scaling the slope from $40$ to $60$ dB.} of $0.31$ and $0.33$ for $\delta = 0.05$ and $0.15$ respectively  ($\bar{d}^{\ast}_{\mathrm{RS}} = 0.333$).
RS schemes give significant performance gains over their NoRS counterparts for the entire SNR range,
with rate gains exceeding $20\%$ and $60 \%$ for $\delta = 0.05$ and $\delta = 0.15$ respectively at high SNRs.

Results for scaling CSIT errors are given in Fig. \ref{Fig_Rate_2}.
The CSIT quality of user-1 remains fixed with $\delta_{1} = \delta$,
while errors for user-2 and user-3 decay with SNR such that $\alpha_{2},\alpha_{3} = 0.5$
and $\beta_{2},\beta_{3} = 10\delta^{2}$,
yielding $\delta_{2},\delta_{3} = \delta \sqrt{10P_{\mathrm{t}}^{-0.5}}$.
Therefore, we have $\delta_{2},\delta_{3} < \delta_{1}$ for SNRs less than $20$ dB, $\delta_{2},\delta_{3} = \delta_{1}$ for $20$ dB SNR, and
$\delta_{2},\delta_{3} > \delta_{1}$ for SNRs greater than $20$ dB.
The general observations made for Fig. \ref{Fig_Rate_1} still hold, with the cs method providing improved performance over the con method, and RS schemes outperforming NoRS schemes.
From a DoF perspective, the cs schemes perform almost as predicted in Theorem \ref{Theorem_RS_NoRS_Max_Min_DoF}: $\bar{d}^{\ast} = 0.25$ and $\bar{d}^{\ast}_{\mathrm{RS}} = 0.5$.
For $\delta = 0.05$ and $0.15$ respectively, NoRS-cs achieves DoFs of $0.26$ and $0.24$, where
RS-cs achieves DoFs of $0.53$ and $0.47$.
On the other hand, NoRS-con and RS-con fail to achieve the corresponding DoFs due to self-interference.

\begin{figure}
  \hspace{-0.05 in}
  \subfloat[][{Fixed CSIT $(\alpha=0)$}]
  {\label{Fig_Rate_31} \includegraphics[width = 0.25\textwidth]{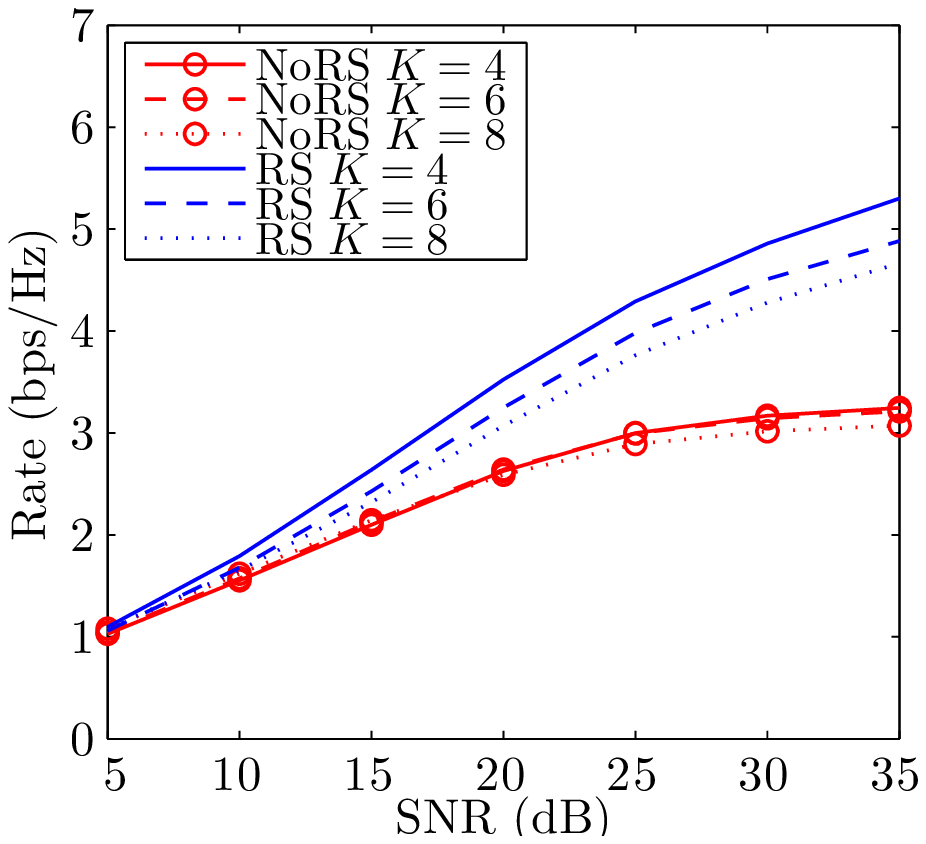}}
  \hspace{-0.2 in}
  \subfloat[][{Scaling CSIT $(\alpha=0.5)$}]
  {\label{Fig_Rate_32} \includegraphics[width = 0.25\textwidth]{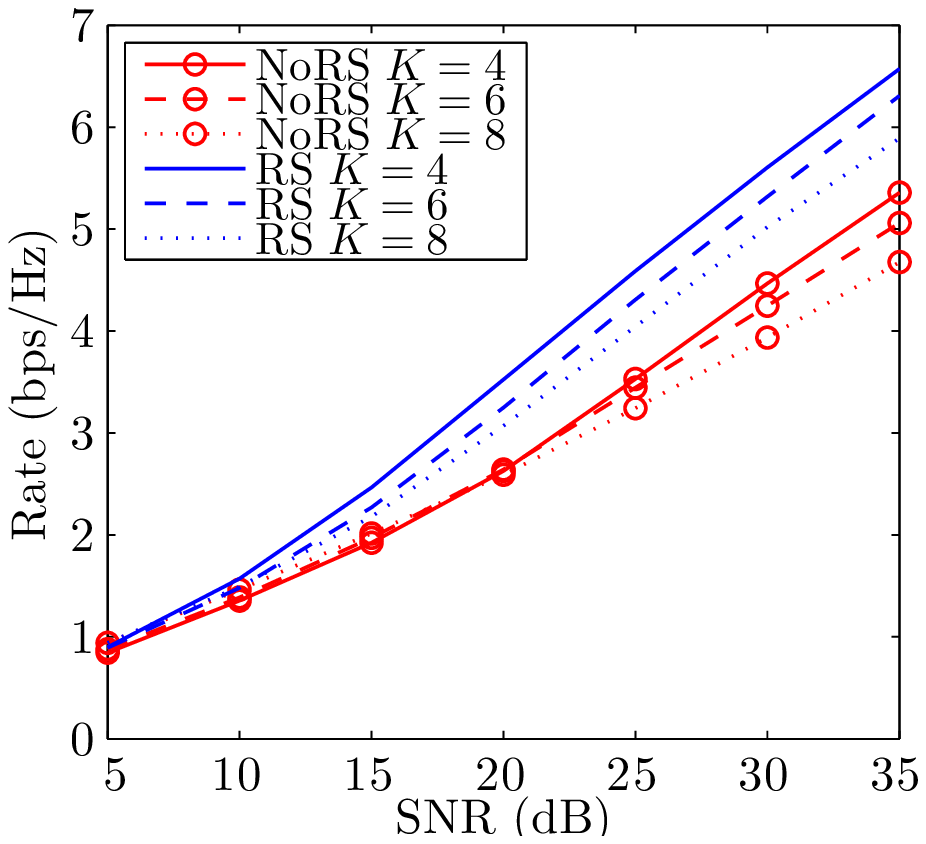}}
  \caption{Cutting-set rate performance for $K,N_{\mathrm{t}} \! = \! 4,6$ and $8$, fixed CSIT: $\delta_1,\ldots,\delta_{K} \! = 0.15$, and scaling CSIT: $\delta_1,\ldots,\delta_{K} \! = 0.15\sqrt{10P_{\mathrm{t}}^{-0.5}}$.}
  \label{Fig_Rate_3}
\vspace{-2.0mm}
\end{figure}

After demonstrating the superiority of the cutting-set method, we examine the RS gains in larger systems with $K,N_{\mathrm{t}} = 4, 6$ and $8$.
The performances of NoRS-cs and RS-cs are given in Fig.
\ref{Fig_Rate_31} for non-scaling CSIT with $\delta_{1},\ldots,\delta_{K} = \delta$,
and Fig. \ref{Fig_Rate_32} for scaling CSIT with $\delta_{1},\ldots,\delta_{K} = \delta\sqrt{10P_{\mathrm{t}}^{-\alpha}}$, where
$\delta = 0.15$ and $\alpha=0.5$.
For a given scheme, the performance generally degrades as the number of users increases. This can be regarded to the increased MU interference in NoRS, in addition to the fact that the common message is shared among more users in RS.
However, the performance gains associated with the RS scheme are still significant.
\subsection{Complexity Comparison}
\label{Subsection_complexity}
Next, we compare the complexities of the conservative method in Algorithm \ref{Algthm_AO_Conservative} and the cutting-set method in Algorithm \ref{Algthm_Cutting_Set}.
We consider $K = N_{\mathrm{t}}$ for simplicity, hence reducing the complexity scaling orders to one parameter.
In each iteration of Algorithm \ref{Algthm_AO_Conservative}, $2K$ equalizers are updated by solving SDPs with a worst-case complexity of $O(K^{3.5})$ each\footnote{Worse-case computational costs of solving standard problems using interior-point methods are given according to \cite[ Lecture 6]{Ben-Tal2001}. The term that accounts for the solution's accuracy is omitted, e.g. \cite{Vucic2009a}.},
while precoders are updated by solving a SDP with a worst-case complexity of $O(K^{8})$.
On the other hand, each cutting-set iteration of Algorithm \ref{Algthm_Cutting_Set} consists of an optimization step and a pessimization step, which are iterative in their own rights.
Each optimization-iteration involves updating the precoders by solving a convex problem with a number of quadratic constraints that grows with the outer (cutting-set) iteration. For the $i$th cutting-set iteration, the number of WMSE constraints cannot exceed $2iK$, and updating the precoders in each inner (optimization) iteration can be formulated as a Second Order Cone Program (SOCP) with a worst-case complexity of $O(i^{1.5}K^{7.5})$.
On the other hand, each pessimization-iteration involves solving $2K$ SDPs with a cost of $O(K^{6.5})$ each.

\begin{figure}
  \hspace{-0.05 in}
  \subfloat[][{$\delta=0.05$}]
  {\label{Fig_run_time_1} \includegraphics[width = 0.25\textwidth]{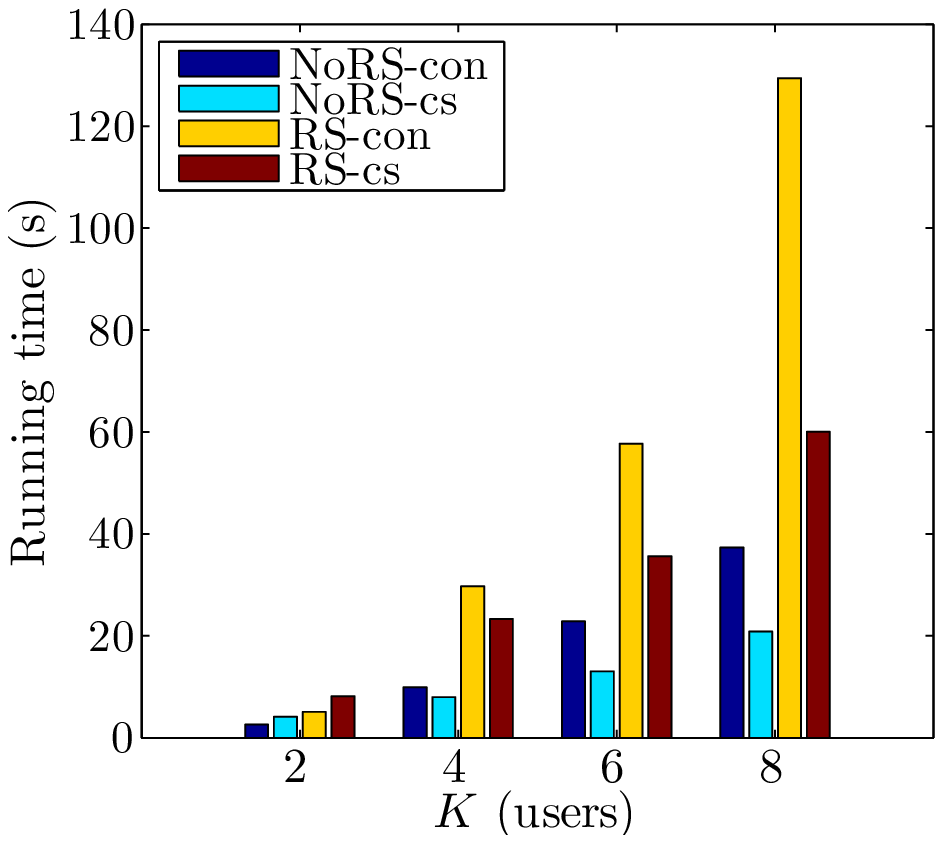}}
  \hspace{-0.2 in}
  \subfloat[][{$\delta=0.15$}]
  {\label{Fig_run_time_2} \includegraphics[width = 0.25\textwidth]{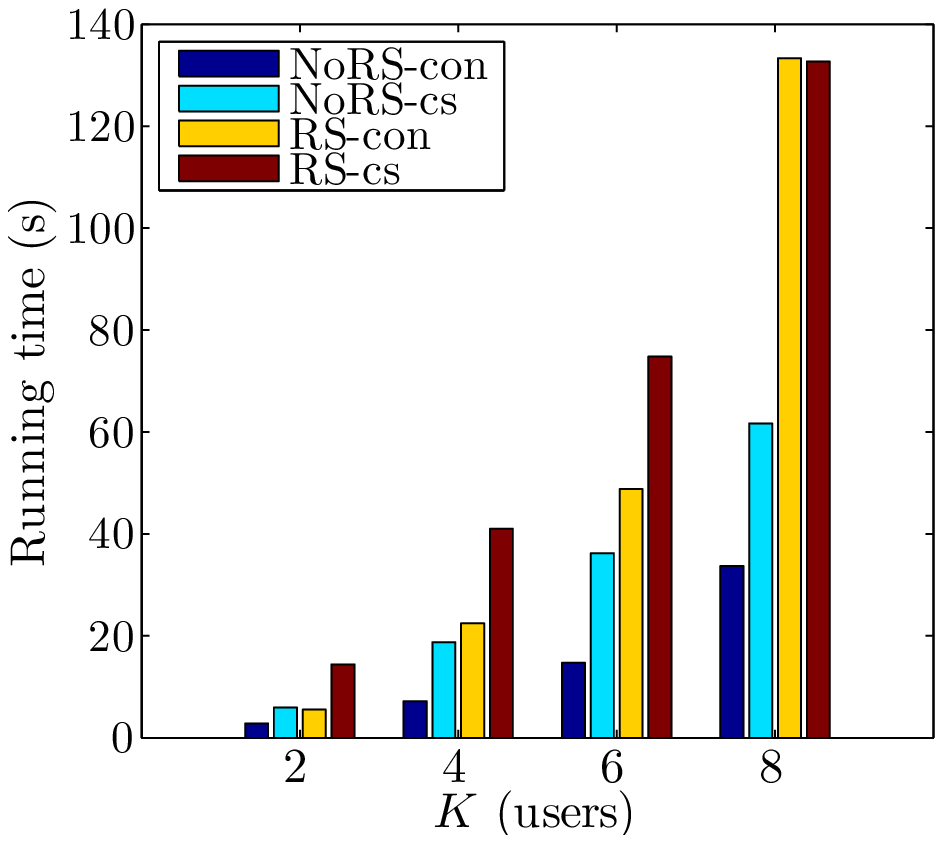}}
  \caption{Average run-time of NoRS and RS, conservative and cutting-set methods for $K,N_{\mathrm{t}} \! = \! 2,4,6$ and $8$, $\mathrm{SNR} \! = \! 20$ dB, and $\delta_1,\ldots,\delta_{K} \! = \! \delta$. }
  \label{Fig_run_time}
\end{figure}

Due to the iterative nature of the two algorithms, in addition to the nested structure of Algorithm \ref{Algthm_Cutting_Set}, a rigorous analytic complexity comparison is not possible.
Alternatively, we evaluate their average running times using MATLAB on a computer equipped with an Intel Core i7-3770 @3.4GHz processor and 8.00 GB of RAM.
Fig. \ref{Fig_run_time} shows the average running times (over $100$ realizations) of the different schemes versus the number of users/antennas at $20$ dB SNR.
For a given method (con or cs), RS has longer running times than NoRS due to the higher number of variables involved.
The con method (NoRS and RS) is hardly influenced by the level of CSIT uncertainty, exhibiting a slight increase in running times for larger $\delta$ due to the increased involvement of the common message, which influences the convergence of the AO algorithm.
On the other hand, the cs method (NoRS and RS) is more influenced by the degree of uncertainty, exhibiting a faster increase in running times with $K$ for higher $\delta$.
This appears to be due to the higher number of pessimization steps required to sample larger uncertainty regions, resulting in an increased number of cutting-set iterations and a growing complexity of the optimization step.
It should be highlighted that in feedback systems, channel quantization codebooks are predetermined and known to the BS.
Hence, corresponding precoders can be calculated beforehand, and relatively long running times do not prohibit the real-time application of such algorithms under limited BS processing capabilities.
\subsection{QoS Constrained Power Minimization}
\label{Subsection_power_problem}
%
In this part we consider the inverse power problem, i.e. minimizing the total transmit power under a minimum rate constraint, also known as the QoS problem.
The RS version of this problem with a minimum rate target $\bar{R}_{\mathrm{t}}$ writes as
\begin{equation}
\label{Eq_Problem_P_RS}
\mathcal{P}_{\mathrm{RS}}(\bar{R}_{\mathrm{t}}):
\begin{cases}
       \underset{\bar{\mathbf{c}} ,\mathbf{P} }{\min} & \mathrm{tr}\big(\mathbf{P}\mathbf{P}^{H}\big) \\
       \text{s.t.}  & \bar{R}_{k} + \bar{C}_{k}  \geq  \bar{R}_{\mathrm{t}},\; \forall k \in  \mathcal{K} \\
                    & \bar{R}_{\mathrm{c},k} \geq \sum_{l=1}^{K} \bar{C}_{l}, \; \forall k\in\mathcal{K} \\
                    & \bar{C}_{k} \geq 0, \; \forall k \in \mathcal{K}.
\end{cases}
\end{equation}
On the other hand, the NoRS counterpart is formulated as
\begin{equation}
\label{Eq_Problem_P_NoRS}
\mathcal{P}(\bar{R}):
\begin{cases}
       \underset{\mathbf{P}_{\mathrm{p}} }{\min} & \mathrm{tr}\big(\mathbf{P}_{\mathrm{p}}\mathbf{P}_{\mathrm{p}}^{H}\big) \\
       \text{s.t.}  & \bar{R}_{k} \geq \bar{R}, \; \forall k\in\mathcal{K}.
\end{cases}
\end{equation}
The power problem is solved using the conservative and cutting-set methods described in the previous sections.
While modifying Algorithm \ref{Algthm_AO_Conservative} and Algorithm \ref{Algthm_Cutting_Set} to address \eqref{Eq_Problem_P_RS} and \eqref{Eq_Problem_P_NoRS} is straightforward, it should be noted that an arbitrary initialization of  $\mathbf{P}$ may easily yield an infeasible point which fails to satisfy the rate constraints.
In this case, the AO algorithm fails to produce a feasible solution, making the initialization a very crucial step.
On the other hand, rate optimization problems are easily initialized by picking any precoder that satisfies $\mathrm{tr}\big( \mathbf{P} \mathbf{P}^{H} \big) \leq P_{\mathrm{t}}$.
This is exploited to obtain a feasible $\mathbf{P}$ for the power problems.
First, let us consider RS-con with a rate constraint $\widehat{R}_{\mathrm{t}}$.
$P_{\mathrm{t}}$ is initialized and the rate optimization procedure in Algorithm \ref{Algthm_AO_Conservative}
is performed until we obtain $\widehat{R}_{\mathrm{t}}^{(n)} \geq \widehat{R}_{\mathrm{t}}$.
The corresponding $\mathbf{P}$ is feasible for the power problem since it satisfies the rate constraint.
If $\widehat{R}_{\mathrm{t}}^{(n)}$ converges before satisfying the rate constraint, $P_{\mathrm{t}}$ is increased
until a feasible point is found.
A feasible $\mathbf{P}$ can be obtained using very few iterations if $P_{\mathrm{t}}$ is adjusted properly.
For RS-cs and NoRS-cs, a similar procedure is followed at the beginning of each optimization step, while noting that some rate constraints may not be feasible for NoRS-cs, and hence $P_{\mathrm{t}}$ should not be increased indefinitely.
NoRS-con boils down to the SDP solution in \cite{Vucic2009}, which does not require initialization.

For power optimization, we only consider non-scaling CSIT, i.e. $\delta_{1}^{2},\ldots,\delta_{K}^{2} = O(1)$ and $\alpha_{1},\ldots,\alpha_{K} = 0$.
This is particularly relevant in this scenario where we assume no BS power constraint, and the CSIT quality is not expected to scale with the transmit power variation during the optimization procedure as channel estimation and feedback is carried out prior to the precoder design.
In the simulations, the minimum rate constraint is set to $3.3219$ bps/Hz, which corresponds to a worst-case user SINR of $9$ dB for the NoRS case \cite{Shenouda2007,Vucic2009}.
The four designs are tested under $100$ channel realizations, for $\delta_{1},\delta_{2},\delta_{3} = \delta$, where $\delta \in \{0.01,0.05,0.1,0.15\}$.
%

\begin{figure}
\vspace{-4.0mm}
  \hspace{-0.05 in}
  \subfloat[][{Num. of feasible realizations}]
  {\label{Fig_feasible}\includegraphics[width = 0.25\textwidth]{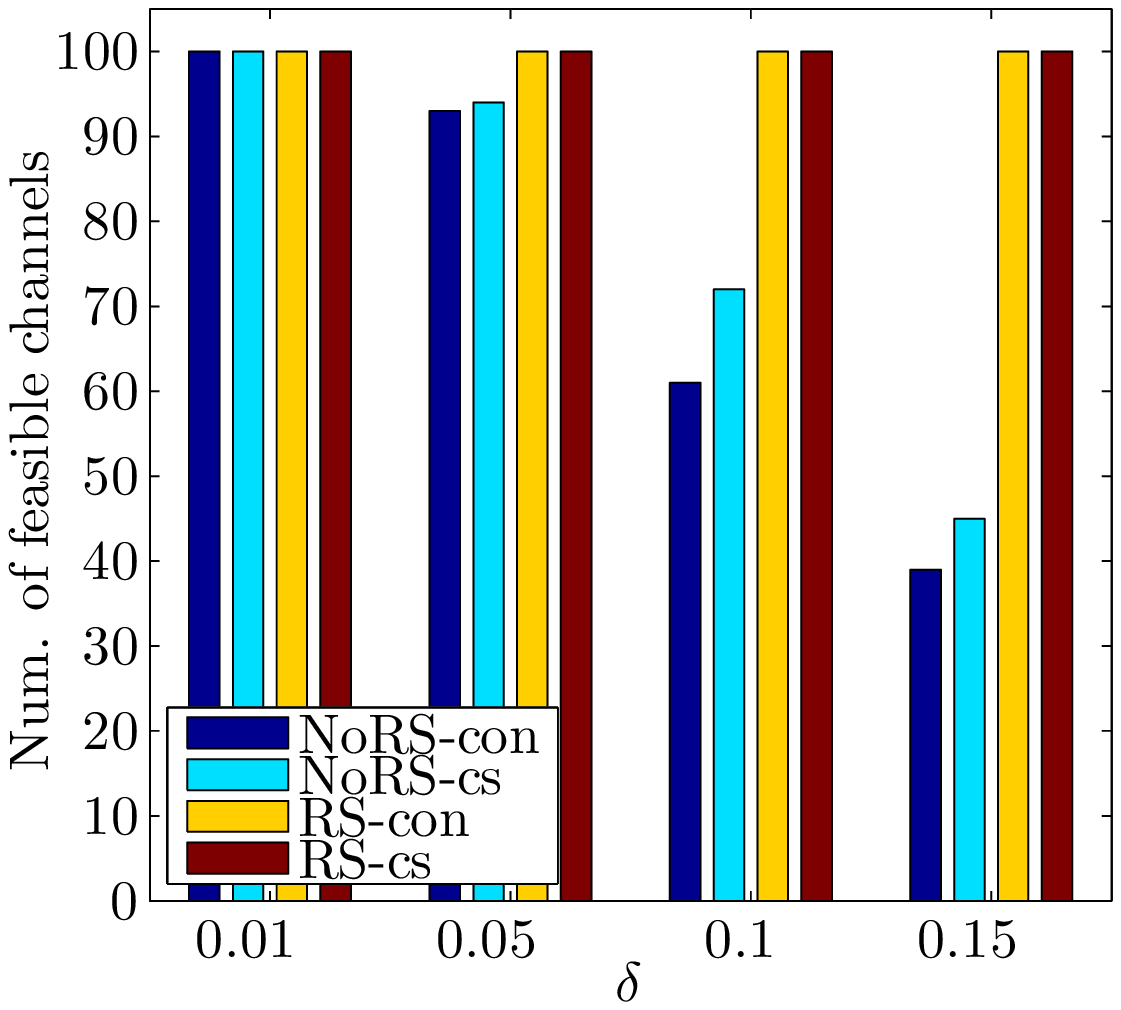}}
  \hspace{-0.1 in}
  \subfloat[][{Transmit power}]
  {\label{Fig_Power}\includegraphics[width = 0.25\textwidth]{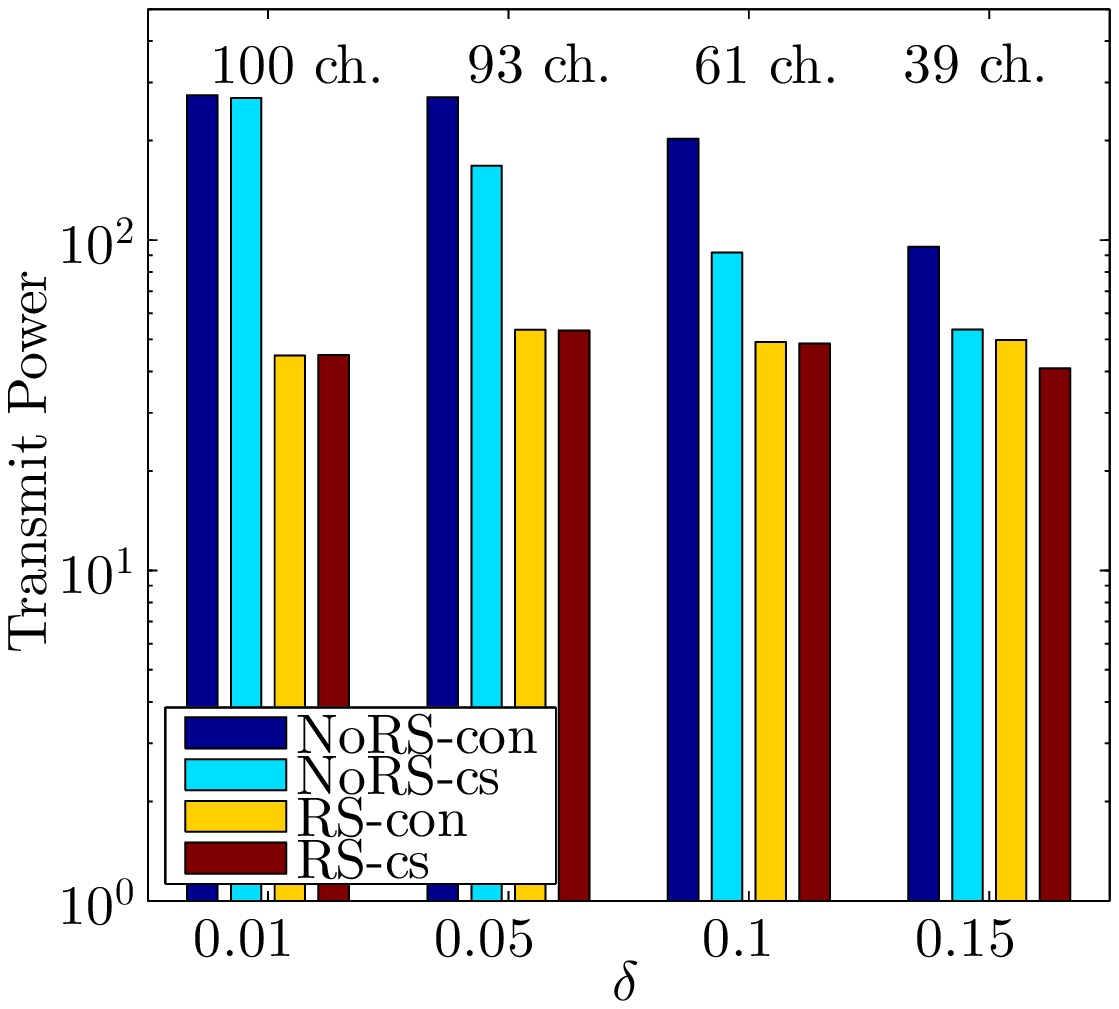}}
  \caption{Power minimization under a QoS constraint of $3.3219$ bps/Hz in a system with $K,N_{\mathrm{t}} = 3$, $\sigma_{\mathrm{n}}^{2} = 1$, and $\delta_{1},\delta_{2},\delta_{3} = \delta$.}
  \label{Fig_QoS}
\vspace{-1.0mm}
\end{figure}
Fig. \ref{Fig_feasible} shows the number of realizations for which the different designs yield a feasible solution.
For the NoRS schemes, the number of feasible channels decreases as the CSIT uncertainty increases.
NoRS-cs outperforms NoRS-con in this regards due to the latter's employment of conservative approximations.
The RS schemes yield feasible solutions for all realizations, with an improvement exceeding $100\%$ compared to NoRS schemes at
$\delta = 0.15$.
This is explained as follows: the rate and the power problems are monotonically non-decreasing in their arguments, and are related such that $\mathcal{R}\big( \mathcal{P}(\bar{R}) \big) = \bar{R}$ and
$\mathcal{R}_{\mathrm{RS}}\big( \mathcal{P}_{\mathrm{RS}}(\bar{R}_{\mathrm{t}}) \big) = \bar{R}_{\mathrm{t}}$, which can be shown by contradiction and power scaling \cite{Wiesel2006,Sidiropoulos2006}.
From the monotonicity of $\mathcal{R}( P_{\mathrm{t}} )$ and Theorem \ref{Theorem_RS_NoRS_Max_Min_DoF}, it follows that under non-scaling CSIT qualities, $\mathcal{R}\big( P_{\mathrm{t}} \big)$ converges to a finite maximum value as $P_{\mathrm{t}} \rightarrow \infty$.
The monotonicity of $\mathcal{P}(\bar{R})$ dictates that this value is the maximum feasible rate.
On the other hand, $\mathcal{R}_{\mathrm{RS}}( P_{\mathrm{t}} )$ does not converge. Therefore, any finite rate is feasible for $\mathcal{P}_{\mathrm{RS}}(\bar{R}_{\mathrm{t}})$, which is always guaranteed by the cutting-set method.
This can also be explained by noting the QoS multicast problem \cite{Sidiropoulos2006}, which is always feasible, is in fact a subproblem of \eqref{Eq_Problem_P_RS}.

Fig. \ref{Fig_Power} shows the total transmit powers averaged over realizations which are feasible for all designs,
i.e. the intersection of the three feasible sets for a given $\delta$.
It can be seen that RS schemes are more efficient in terms of total transmit power compared to NoRS designs.
Intuitively, we expect this contrast to increase with $\delta$
(by reversing the observations in Fig. \ref{Fig_Rate_1}).
This holds if infeasible realizations are assigned infinitely large transmit powers.
However, since more realizations are omitted for increased $\delta$, the powers obtained in Fig. \ref{Fig_feasible}
for a larger $\delta$ are in fact averaged over very well conditioned channels.
%
\section{Conclusion}
\label{Section_Conclusion}
The classical robust optimization problem of achieving max-min fairness in a MU-MISO system with bounded CSIT errors was addressed using an unconventional RS transmission strategy.
We analytically proved that the proposed RS strategy outperforms the conventional NoRS strategy in the interference limited regime.
Although a solution for the RS design problem can be obtained using the conservative WMMSE approach in \cite{Tajer2011}, we demonstrated the limitations of such approach through deriving an upper-bound on the resulting conservative performance.
This upper-bound explains the saturating performances observed in \cite{Tajer2011} and \cite{Joudeh2016} which contradict the predictions from the DoF analysis.
Alternatively, we proposed a non-conservative algorithm based on the cutting-set method, and proved its convergence to the set of KKT points of the non-convex optimization problem.
The superiority of the proposed algorithm and the gains of RS were demonstrated through simulations.
The approach was also extended to solve the QoS problem, where it was shown that RS eliminates the feasibility issue arising in NoRS designs.     
A less pronounced yet highly important contribution of this paper is that it invites a rethinking of robust designs in other interference-limited scenarios in the light of the RS strategy, for example: multi-cell transmission \cite{Tajer2011}, cognitive radio beamforming \cite{Gharavol2010}, and energy efficient beamforming \cite{Xu2015}, to name a few.
\appendices
\section{Proof of Theorem \ref{Theorem_RS_NoRS_Max_Min_DoF}}
\label{Appendix_Proof_Theorem_RS_NoRS_Max_Min_DoF}
%
The following lemmas are instrumental to the proof. First, let us define the function $(x)^{+} \triangleq \max{\{x,0\}}$, where $x \in \mathbb{R}$.
\vspace{-1.0mm}
\newtheorem{Lemma_inner_prod_UB_LB}[Lemma_Counter]{Lemma}
\begin{Lemma_inner_prod_UB_LB}\label{Lemma_inner_prod_UB_LB}
\textnormal{\cite[Lemma 1]{Tajer2011}}
\textnormal{Given the ball uncertainty model
and for any $\mathbf{p} \in \mathbb{C}^{N_{t}}$, we have
\vspace{-2.0mm}
\begin{align}
\nonumber
\max_{\mathbf{h}_{k} \in \mathbb{H}_{k} }|\mathbf{h}_{k}^{H} \mathbf{p}| & = | \widehat{\mathbf{h}}_{k}^{H} \mathbf{p} | + \delta_{k} \| \mathbf{p} \|\\
\nonumber
\min_{\mathbf{h}_{k} \in \mathbb{H}_{k} } |\mathbf{h}_{k}^{H} \mathbf{p}| & = \big( | \widehat{\mathbf{h}}_{k}^{H} \mathbf{p} | - \delta_{k} \| \mathbf{p} \| \big)^{+}.
\end{align}}
\end{Lemma_inner_prod_UB_LB}
\vspace{-1.0mm}
\newtheorem{Lemma_achievable_DoF}[Lemma_Counter]{Lemma}
\begin{Lemma_achievable_DoF}\label{Lemma_achievable_DoF}
\textnormal{There exists a feasible RS precoding scheme that achieves the DoF
\vspace{-2.0mm}
\begin{equation}
\hat{d}_{\mathrm{c}}
= 1 - \bar{a}
\quad \text{and} \quad
\hat{d}_{k}
= \min\{(\alpha_{k} + a_{k} - \bar{a}_{k})^{+},a_{k}\}
\end{equation}
for all ${a}_{k} \in [0,1]$, $\bar{a} \triangleq \max_{j}\{a_{j}\}_{j=1}^{K}$ and $\bar{a}_{k} \triangleq \max_{j}\{a_{j}\}_{j \neq k}$.}
\end{Lemma_achievable_DoF}
\begin{proof}[Proof]
We consider a ZF-BF design based on the imperfect estimate $\widehat{\mathbf{H}}$ for the private precoders such that
$\mathbf{P}_{\mathrm{p}} = \big( \widehat{\mathbf{H}}^{H}\big)^{\dagger} \mathbf{B}$, where
$\mathbf{B} \triangleq \mathrm{diag}\big( \sqrt{q_{1}/b_{1}},\ldots,\sqrt{q_{K}/b_{K}} \big)$,
and $b_{1},\ldots,b_{K}$ are constants that normalize the columns of $\big( \widehat{\mathbf{H}}^{H}\big)^{\dagger}$.
The existence of such solution is guaranteed by Assumption \ref{Assumption_Bounded_Channel}.
The common precoder is given as $\mathbf{p}_{\mathrm{c}} = \sqrt{q_{\mathrm{c}}} \mathbf{e}_{1}$, where $a_{\mathrm{c}}$ is set to $1$.
We define the worst-case SINRs as
$\bar{\gamma}_{\mathrm{c},k} \triangleq  \underset{\mathbf{h}_{k} \in
\mathbb{H}_{k}}{\min} \gamma_{\mathrm{c},k}\big(\mathbf{h}_{k}\big)$
and
$\bar{\gamma}_{k} \triangleq  \underset{\mathbf{h}_{k} \in
\mathbb{H}_{k}}{\min} \gamma_{k}\big(\mathbf{h}_{k}\big)$\footnote{Worst-case channels are equivalently obtained using the rates or SINRs}. %
By applying the described scheme, $\bar{\gamma}_{k}$ is lower-bounded as
\begin{equation}
\label{Eq_SINR_k_LB}
\bar{\gamma}_{k} \geq \frac{q_{k}(\sqrt{1/b_{k}} - \delta_{k})^{2} }{\displaystyle{\sum_{i \neq k}} \big| \widetilde{\mathbf{h}}_{k}^{H}\mathbf{p}_{i} \big|^{2} + \sigma_{\mathrm{n}}^{2} } \\
\geq
\frac{q_{k}(\sqrt{1/b_{k}} - \delta_{k})^{2}}{\delta_{k}^{2}\displaystyle{\sum_{i \neq k}} q_{i}   + \sigma_{\mathrm{n}}^{2} }.
\end{equation}
The left inequality in \eqref{Eq_SINR_k_LB} follows from Lemma \ref{Lemma_inner_prod_UB_LB} and the assumption that
$|\widehat{\mathbf{h}}_{k}^{H} \mathbf{p}_{k}|^{2} > \delta_{k}^{2}q_{k}$, i.e. small error \cite{Vorobyov2003}.
The right inequality is obtained from applying the Cauchy-Schwarz inequality and
$\|\widetilde{\mathbf{h}}_{k}\|^{2} \leq \delta_{k}^{2}$ to the denominator.
The numerator scales as $O\big(P_{\mathrm{t}}^{a_{k}}\big)$, while the denominator scales as
$O\big(P_{\mathrm{t}}^{(\bar{a}_{k} - \alpha_{k})^{+}}\big)$.
It follows from \eqref{Eq_R} and \eqref{Eq_Wors_Case_DoF} that
$\bar{d}_{k} \geq \min\{(\alpha_{k} + a_{k} - \bar{a}_{k})^{+},a_{k}\}$.
For $\bar{\gamma}_{\mathrm{c},k}$, we write
\begin{equation}
\label{Eq_SINR_c_LB}
\bar{\gamma}_{\mathrm{c},k} \geq \frac{q_{\mathrm{c}} |h_{k,1}|^{2}}
{\|\mathbf{h}_{k}\|^{2}\sum_{k = 1}^{K} q_{k} + \sigma_{\mathrm{n}}^{2} } \\
= O\big( P_{\mathrm{t}}^{(1-\bar{a})} \big)
\end{equation}
where the Cauchy-Schwarz inequality is applied to the denominator, and both $|h_{k,1}|^{2}$ and $\|\mathbf{h}_{k}\|^{2}$ scale as $O(1)$ from Assumption \ref{Assumption_Bounded_Channel}.
From \eqref{Eq_R} and \eqref{Eq_Wors_Case_DoF}, we have $\bar{d}_{c} \geq 1 - \bar{a}$.
\end{proof}
%
\begin{proof}[Proof of Theorem \ref{Theorem_RS_NoRS_Max_Min_DoF}]
To characterize the optimum DoF performance, we define the optimum precoding schemes for  \eqref{Eq_Problem_R_RS} and \eqref{Eq_Problem_R_NoRS} as $\big\{\mathbf{P}^{\ast}(P_{\mathrm{t}})\big\}_{P_{\mathrm{t}}}$ and
$\big\{\mathbf{P}_{\mathrm{p}}^{\ast}(P_{\mathrm{t}})\big\}_{P_{\mathrm{t}}}$ respectively, where the corresponding powers and exponents are
denoted by $q_{\mathrm{c}}^{\ast}$, $q_{k}^{\ast}$, $a_{\mathrm{c}}^{\ast}$ and $a_{k}^{\ast}$.
\subsubsection{Proof of \eqref{Eq_Max_Min_DoF_NoRS}}
We start by showing that for any given precoding scheme with a given power allocation, the achievable private DoF in Lemma \ref{Lemma_achievable_DoF} cannot be exceeded, i.e.
\begin{equation}
\label{Eq_dk_wc_UB}
\bar{d}_{k} \leq \min{\{(\alpha_{k} + a_{k} - \bar{a}_{k})^{+},a_{k}\}}.
\end{equation}
The worst-case SINR is upper-bounded  as
$\bar{\gamma}_{k} \leq \gamma_{k}(\mathbf{h}_{k} )$, where $\mathbf{h}_{k} \in \mathbb{H}_{k}$.
$\mathbf{h}_{k} $ is selected such that the $l$th user's interference term is maximized in accordance with Lemma \ref{Lemma_inner_prod_UB_LB},
i.e.
$\big|\widehat{\mathbf{h}}_{k}^{H}\mathbf{p}_{l}+\widetilde{\mathbf{h}}_{k}^{H}\mathbf{p}_{l}\big| = \big|\widehat{\mathbf{h}}_{k}^{H}\mathbf{p}_{l}\big| + \delta_{k} \| \mathbf{p}_{l} \|$,
where $l$ is chosen such that $a_{l} = \bar{a}_{k} \triangleq \max{\{a_{j}\}}_{j\neq k} $.
As a result, we obtain the upper-bound
\begin{align}
\nonumber
\bar{\gamma}_{k}
& \leq \frac{\big|\mathbf{h}_{k}^{H}\mathbf{p}_{k} \big|^{2}}
{\big(  |\widehat{\mathbf{h}}_{k}^{H}\mathbf{p}_{l}|  \! + \!
\delta_{k} \| \mathbf{p}_{l}\|  \big)^2 \!  \! + \! \!  \displaystyle{\sum_{i \neq k,l}} \big|\widehat{\mathbf{h}}_{k}^{H}\mathbf{p}_{i}  \! + \!  \widetilde{\mathbf{h}}_{k}^{H}\mathbf{p}_{i} \big|^{2} \! + \! \sigma_{\mathrm{n}}^{2}
} \\
\label{Eq_wc_SINR_UB}
& \leq
\frac{ \| \mathbf{h}_{k} \|^{2} q_{k} }
{\delta_{k}^{2} q_{l} +  \sigma_{\mathrm{n}}^{2}}.
\end{align}
where \eqref{Eq_wc_SINR_UB} follows from applying the Cauchy-Schwarz inequality and discarding non-negative interference terms.
From Assumption \ref{Assumption_Bounded_Channel}, it is evident that \eqref{Eq_wc_SINR_UB} scales as the lower-bound in \eqref{Eq_SINR_k_LB}, from which \eqref{Eq_dk_wc_UB} directly follows.
The optimum DoF satisfies $\bar{d}^{\ast} \leq \bar{d}_{k}^{\ast}, \forall k\in \mathcal{K}$, where $\bar{d}_{k}^{\ast}$ is the $k$th user's DoF at optimality.
From \eqref{Eq_dk_wc_UB}, we write
\begin{equation}
\label{Eq_Max_Min_DoF_UB_1}
\bar{d}^{\ast} \leq  \min_{k} \big\{ \min\{\alpha_{k} + a_{k}^{\ast} - \bar{a}_{k}^{\ast},a_{k}^{\ast}\}  \big\}_{k=1}^{K}
\end{equation}
where $(\cdot)^{+}$ is omitted by assuming that
$(\alpha_{k} + a_{k}^{\ast} - \bar{a}_{k}^{\ast}) \geq 0$.
This assumption is valid as $(\alpha_{k} + a_{k}^{\ast} - \bar{a}_{k}^{\ast}) < 0$ yields $\bar{d}^{\ast} = 0$, which is maintained if $a_{k}^{\ast}$ is increased to $\bar{a}_{k}^{\ast} - \alpha_{k}$.
On the other hand, $\bar{d}^{\ast}  > 0$ is only obtained when $(\alpha_{k} + a_{k}^{\ast} - \bar{a}_{k}^{\ast}) > 0$.
\eqref{Eq_Max_Min_DoF_UB_1} is further upper-bounded as
\begin{align}
\label{Eq_Max_Min_DoF_UB_2}
\bar{d}^{\ast} & \leq \! \frac{ \min\{\alpha_{1} \! + \! a_{1}^{\ast} \! - \! \bar{a}_{1}^{\ast},a_{1}^{\ast}\}  +
\min\{\alpha_{2} \! + \! a_{2}^{\ast} \! - \! \bar{a}_{2}^{\ast},a_{2}^{\ast}\} }{2} \\
\label{Eq_Max_Min_DoF_UB_3}
& \leq \frac{ \alpha_{1} + a_{1}^{\ast} - \bar{a}_{1}^{\ast} + \alpha_{2} + a_{2}^{\ast} - \bar{a}_{2}^{\ast} }{2} \\
\label{Eq_Max_Min_DoF_UB_4}
& \leq \frac{\alpha_{1} + \alpha_{2}}{2}.
\end{align}
\eqref{Eq_Max_Min_DoF_UB_2} follows from  the fact that $\bar{d}^{\ast}$ is upper-bounded by the average of any two DoFs,
and \eqref{Eq_Max_Min_DoF_UB_3} is obtained by noting that the point-wise minimum is upper-bounded by any element in the set.
\eqref{Eq_Max_Min_DoF_UB_4} follows from $a_{j}^{\ast} \leq \bar{a}_{k}^{\ast}, \; \forall j \neq k$.
From Lemma \ref{Lemma_achievable_DoF}, allocating the private powers such that
$a_{1} = \alpha_{2}$ and $a_{2},\ldots,a_{K} = \frac{\alpha_{1}+\alpha_{2}}{2}$,
we achieve $\bar{d}_{k} \geq \frac{\alpha_{1} + \alpha_{2}}{2}, \; \forall k \in \mathcal{K}$.
%
\subsubsection{Proof of \eqref{Eq_Max_Min_DoF_RS}}
We start this part by showing that
\begin{equation}
\label{Eq_dc_dk_wc_UB}
\bar{d}_{\mathrm{c}} + \bar{d}_{k} \leq \min{\{1+\alpha_{k} - \bar{a}_{k},1\}}.
\end{equation}
This result follows from
\begin{align}
\nonumber
\bar{R}_{\mathrm{c}} + \bar{R}_{k} & \leq \bar{R}_{\mathrm{c},k} +  \bar{R}_{k}  \\
\label{Eq_Rc_Rk_UB_2}
& \leq R_{\mathrm{c},k}( \mathbf{h}_{k} \big) +  R_{k}\big( \mathbf{h}_{k} \big) \\
\label{Eq_Rc_Rk_UB_3}
& = \log_{2}\Big(T_{\mathrm{c},k}\big( \mathbf{h}_{k} \big) \Big) -
\log_{2}\Big(I_{k}\big( \mathbf{h}_{k} \big) \Big) \\
\label{Eq_Rc_Rk_UB_4}
& = \log_{2}(P_{\mathrm{t}}) - (\bar{a}_{k} - \alpha_{k} )^{+}\log_{2}(P_{\mathrm{t}})  + O(1)
\end{align}
where \eqref{Eq_Rc_Rk_UB_2} is obtained using the same $\mathbf{h}_{k}$ employed in \eqref{Eq_wc_SINR_UB},
\eqref{Eq_Rc_Rk_UB_3} follows from the rate definitions, and
\eqref{Eq_Rc_Rk_UB_4} is obtained using means of previous analysis.

The optimum DoF satisfies $\bar{d}^{\ast}_{\mathrm{RS}} \leq \bar{c}_{k}^{\ast} + \bar{d}_{k}^{\ast}, \; \forall k \in \mathcal{K} $, where $\sum_{k=1}^{K}\bar{c}_{k}^{\ast} = \bar{d}_{\mathrm{c}}^{\ast}$ and $\bar{c}_{k}^{\ast} \geq 0$.
An upper-bound is obtained by taking the average of any number of user DoFs.
To obtain a tighter upper-bound, we optimize over the number of averaged users such that
\begin{equation}
\label{Eq_Max_Min_DoF_RS_UB_2}
\bar{d}^{\ast}_{\mathrm{RS}}  \leq \min_{J\in \mathcal{K}}
\frac{\sum_{k=1}^{J} \big( \bar{c}_{k}^{\ast} + \bar{d}_{k}^{\ast} \big) }{J}
\leq \min_{J\in \mathcal{K}}  \frac{\bar{d}_{\mathrm{c}}^{\ast}  + \sum_{k=1}^{J}\bar{d}_{k}^{\ast}}{J}
\end{equation}
where  \eqref{Eq_Max_Min_DoF_RS_UB_2} follows from $ \sum_{k=1}^{J} \bar{c}_{k}^{\ast} \leq \bar{d}_{\mathrm{c}}^{\ast}$.
The argument used to omit $(\cdot)^{+}$ in \eqref{Eq_Max_Min_DoF_UB_1} cannot be directly applied for \eqref{Eq_Max_Min_DoF_RS_UB_2}.
Alternatively, we start by assuming that $(\alpha_{k} + a_{k}^{\ast} - \bar{a}_{k}^{\ast}) \geq 0, \; \forall k \in \{1,\ldots,J\}$, for a given $J$.
For the case where $J$ is an odd number, we write
\begin{align}
\nonumber
\frac{\bar{d}_{\mathrm{c}}^{\ast}  + \sum_{k=1}^{J}\bar{d}_{k}^{\ast}}{J}
 & = \frac{\bar{d}_{\mathrm{c}}^{\ast} + \bar{d}_{J}^{\ast} + \sum_{k=1}^{J-1}\bar{d}_{k}^{\ast}}{J} \\
\label{Eq_d_RS_UB_odd_1}
& \leq  \frac{ 1 + \sum_{k=1}^{J-1} (\alpha_{k} + a_{k}^{\ast} - \bar{a}_{k}^{\ast}) }{J}\\
\label{Eq_d_RS_UB_odd_2}
& \leq  \frac{ 1 + \sum_{k=1}^{J-1}\alpha_{k} }{J}.
\end{align}
\eqref{Eq_d_RS_UB_odd_1} follows from \eqref{Eq_dc_dk_wc_UB} and \eqref{Eq_dk_wc_UB}, where the elements $1$ and $(\alpha_{k} + a_{k}^{\ast} - \bar{a}_{k}^{\ast})$ are picked to upper-bound $\bar{d}_{\mathrm{c}}^{\ast} + \bar{d}_{J}^{\ast}$ and $\bar{d}_{k}^{\ast}$ respectively.
\eqref{Eq_d_RS_UB_odd_2} is obtained by writing the sum in \eqref{Eq_d_RS_UB_odd_1} as a sum of pairs, and using the approach in
\eqref{Eq_Max_Min_DoF_UB_4}.
For the case where $J$ is an even number, we write
\begin{align}
\nonumber
\frac{\bar{d}_{\mathrm{c}}^{\ast}  \! + \! \sum_{j=1}^{J} \! \bar{d}_{j}^{\ast}}{J}
& \! = \! \frac{\bar{d}_{\mathrm{c}}^{\ast} + \bar{d}_{1}^{\ast} + \bar{d}_{J}^{\ast} + \sum_{k=2}^{J-1}\bar{d}_{k}^{\ast}}{J} \\
\label{Eq_d_RS_UB_even_1}
& \! \leq \!  \frac{ 1 \! + \! \alpha_{1} \! - \! \bar{a}_{1}^{\ast} \! + \! a^{\ast}_{J} \! + \! \sum_{k=2}^{J-1}  (  \alpha_{k} \! +  \! a_{k}^{\ast} \! - \! \bar{a}_{k}^{\ast}  ) }{J}\\
\label{Eq_d_RS_UB_even_2}
& \leq  \frac{ 1 + \sum_{k=1}^{J-1}\alpha_{k} }{J}.
\end{align}
In \eqref{Eq_d_RS_UB_even_1},
$1 + \alpha_{1} - \bar{a}_{1}^{\ast}$ and $a^{\ast}_{J}$ are chosen to upper-bound $\bar{d}_{\mathrm{c}}^{\ast} + \bar{d}_{1}^{\ast}$
and $\bar{d}_{J}^{\ast}$ respectively.
\eqref{Eq_d_RS_UB_even_2} is obtained from $\bar{a}_{1}^{\ast} \geq a_{J}^{\ast}$ and the approach in \eqref{Eq_d_RS_UB_odd_2}.
If we assume that a given $(\alpha_{k} + a^{\ast}_{k} - \bar{a}_{k}^{\ast}) < 0$ for a subset of $\{1,\ldots,J\}$, and hence $\bar{d}_{j}^{\ast} = 0$, we cannot exceed \eqref{Eq_d_RS_UB_odd_2} and \eqref{Eq_d_RS_UB_even_2}.
Combining this with \eqref{Eq_Max_Min_DoF_RS_UB_2}, we obtain
\begin{equation}
\label{Eq_Max_Min_DoF_RS_UB_3}
\bar{d}^{\ast}_{\mathrm{RS}}  \leq \min_{J\in \{2,\ldots,K \}}
\frac{ 1 + \sum_{j=1}^{J-1} \alpha_{j} }{J}.
\end{equation}
where $J=1$ has been omitted.
Next, we show that this upper-bound is achievable by a feasible precoding scheme.
From Lemma \ref{Lemma_achievable_DoF}, allocating the powers such that $a_{k} = \bar{a}$ for all $k$, we achieve DoFs $\bar{d}_{k}$ and $\bar{d}_{\mathrm{c}}$ of $\min{ \{\alpha_{k},\bar{a}\}}$ and $1-\bar{a}$ respectively.
We show that there exists $\bar{a} \in [0,1]$ and feasible $\{\bar{c}_{k}\}_{k=1}^{K}$ such that
$\bar{c}_{k} + \min{\{\alpha_{k},\bar{a}\}}$ achieves the upper-bound in \eqref{Eq_Max_Min_DoF_RS_UB_3}.

For a given $J$, the corresponding upper-bound $\frac{ 1 + \sum_{j=1}^{J-1} \alpha_{j} }{J}$ is denoted by $\bar{d}_{\mathrm{RS}}^{\mathrm{UB}}(J)$.
Let $J^{\ast}$ be the argument of the minimization in \eqref{Eq_Max_Min_DoF_RS_UB_3}, i.e.
$\bar{d}_{\mathrm{RS}}^{\mathrm{UB}}(J^{\ast}) \leq \bar{d}_{\mathrm{RS}}^{\mathrm{UB}}(J)$.  If $J^{\ast} < K$,
\begin{equation}
\label{Eq_d_rs_ub_bounds}
  \alpha_{J^{\ast} - 1} \leq \bar{d}_{\mathrm{RS}}^{\mathrm{UB}}(J^{\ast}) \leq \alpha_{J^{\ast}}
\end{equation}
which is shown in the following. First, we note that
\begin{align}
\label{Eq_d_rs_ub_J_1}
\bar{d}_{\mathrm{RS}}^{\mathrm{UB}}(J+1) & = \frac{J\bar{d}_{\mathrm{RS}}^{\mathrm{UB}}(J) + \alpha_{J}}{J+1} \\
\label{Eq_d_rs_ub_J_2}
\bar{d}_{\mathrm{RS}}^{\mathrm{UB}}(J-1) & = \frac{J\bar{d}_{\mathrm{RS}}^{\mathrm{UB}}(J) - \alpha_{J-1}}{J-1}.
\end{align}
Since $\bar{d}_{\mathrm{RS}}^{\mathrm{UB}}(J^{\ast}) \leq \bar{d}_{\mathrm{RS}}^{\mathrm{UB}}(J^{\ast}+1)$ and $\bar{d}_{\mathrm{RS}}^{\mathrm{UB}}(J^{\ast}) \leq \bar{d}_{\mathrm{RS}}^{\mathrm{UB}}(J^{\ast}-1)$,
the right and left inequalities in \eqref{Eq_d_rs_ub_bounds} follow from \eqref{Eq_d_rs_ub_J_1} and \eqref{Eq_d_rs_ub_J_2} respectively,
as the average increases by including $\alpha_{J}$ in  \eqref{Eq_d_rs_ub_J_1} and excluding $\alpha_{J-1}$ in  \eqref{Eq_d_rs_ub_J_2}.
For $J^{\ast} = K$, the right inequality in \eqref{Eq_d_rs_ub_bounds} does not necessarily hold, but the left inequality always holds.
Hence, we have one of the two following cases.
\begin{enumerate}
\item \textbf{$J^{\ast} < K$ or $J^{\ast} = K$ and \eqref{Eq_d_rs_ub_bounds} holds:}
for this case, we set $\bar{a} = \bar{d}_{\mathrm{RS}}^{\mathrm{UB}}(J^{\ast})$. We obtain DoFs of
$\bar{d}_{k}  =  \alpha_{k},\forall k < J^{\ast}$,
$\bar{d}_{k} = \bar{d}_{\mathrm{RS}}^{\mathrm{UB}}(J^{\ast}), \forall k \geq J^{\ast}$,
and $\bar{d}_{\mathrm{c}} = 1 - \bar{d}_{\mathrm{RS}}^{\mathrm{UB}}(J^{\ast})$.
The common DoF is split such that $\bar{c}_{k} = \bar{d}_{\mathrm{RS}}^{\mathrm{UB}}(J^{\ast}) - \alpha_{k}, \forall k < J^{\ast}$, and
$\bar{c}_{k} = 0, \forall k \geq J^{\ast}$.
The left inequality in \eqref{Eq_d_rs_ub_bounds} guarantees that $\bar{c}_{k} \geq 0$,
while we can see that
$\sum_{k=1}^{J^{\ast} - 1}\bar{c}_{k} = (J^{\ast} - 1)\bar{d}_{\mathrm{RS}}^{\mathrm{UB}}(J^{\ast}) -  \sum_{k=1}^{J^{\ast} - 1}\alpha_{k} = \bar{d}_{\mathrm{c}} $.
\item \textbf{$J^{\ast} = K$  and $\bar{d}_{\mathrm{RS}}^{\mathrm{UB}}(K)\geq \alpha_{k}$ holds $\forall k \in \mathcal{K}$:}
we set $\bar{a} =\alpha_{K}$ obtaining DoFs of $\bar{d}_{k} = \alpha_{k}$ and
$\bar{d}_{\mathrm{c}} = 1 - \alpha_{K}$. The common DoF is split as: $\bar{c}_{k} = \bar{d}_{\mathrm{RS}}^{\mathrm{UB}}(K) - \alpha_{K}$,
which are non-negative and satisfy $\sum_{k=1}^{K}\bar{c}_{k} = \bar{d}_{\mathrm{c}}$.
\end{enumerate}
This completes the proof.
\end{proof}
\section{Proof of Lemma \ref{Lemma_R_hat_UB}}
\label{Appendix_Proof_Lemma_R_hat_UB}
For any pair $(\widehat{g}_{\mathrm{c},k}, \widehat{u}_{\mathrm{c},k})$, independent of the actual channel realization, and a given estimate $\widehat{\mathbf{h}}_{k}$, averaging the MSEs in \eqref{Eq_MSE} over the error distribution in Section \ref{Subsection_Conservative_Limitations} yields
\begin{subequations}
\label{Eq_average_MSE}
\begin{align}
  \widehat{\varepsilon}_{\mathrm{c},k}(\widehat{g}_{\mathrm{c},k}, \widehat{u}_{\mathrm{c},k}) & = |\widehat{g}_{\mathrm{c},k}|^{2} \widehat{T}_{\mathrm{c},k} -2\Re \big\{\widehat{g}_{\mathrm{c},k}\widehat{\mathbf{h}}_{k}^{H}\mathbf{p}_{\mathrm{c}}\big\}+1 \\
  \widehat{\varepsilon}_{k}(\widehat{g}_{\mathrm{c},k}, \widehat{u}_{\mathrm{c},k}) & =  |\widehat{g}_{k}|^{2} \widehat{T}_{k}-2\Re \big\{\widehat{g}_{k}\widehat{\mathbf{h}}_{k}^{H}\mathbf{p}_{k}\big\}+1.
\end{align}
\end{subequations}
The optimum equalizers for \eqref{Eq_average_MSE} are given by $\widehat{g}_{\mathrm{c},k}^{\mathrm{MMSE}} = \mathbf{p}_{\mathrm{c}}^{H}\widehat{\mathbf{h}}_{k}\widehat{T}_{\mathrm{c},k}^{-1}$ and
$\widehat{g}_{k}^{\mathrm{MMSE}} = \mathbf{p}_{k}^{H}\widehat{\mathbf{h}}_{k}\widehat{T}_{k}^{-1}$.
The corresponding average MMSEs are given by $ \widehat{\varepsilon}_{\mathrm{c},k}^{\mathrm{MMSE}} = 1 - \widehat{T}_{\mathrm{c},k}^{-1}|\mathbf{p}_{\mathrm{c}}^{H}\widehat{\mathbf{h}}_{k}|^{2}$ and
$\widehat{\varepsilon}_{k}^{\mathrm{MMSE}} =  1 - \widehat{T}_{k}^{-1}|\mathbf{p}_{k}^{H}\widehat{\mathbf{h}}_{k}|^{2}$.
In a similar manner, averaging the WMSEs in \eqref{Eq_A_WMSEs} yields
\begin{subequations}
\label{Eq_average_WMSE}
\begin{align}
\widehat{\xi}_{\mathrm{c},k}\big( \widehat{g}_{\mathrm{c},k}, \widehat{u}_{\mathrm{c},k} \big) & =
\widehat{u}_{\mathrm{c},k} \widehat{\varepsilon}_{\mathrm{c},k}( \widehat{g}_{\mathrm{c},k}, \widehat{u}_{\mathrm{c},k} \big)  -  \log_{2} (  \widehat{u}_{\mathrm{c},k} ) \\
\widehat{\xi}_{k}\big( \widehat{g}_{k}, \widehat{u}_{k} \big) & =
\widehat{u}_{k} \widehat{\varepsilon}_{k}( \widehat{g}_{k}, \widehat{u}_{k} \big)  -  \log_{2} (  \widehat{u}_{k} ).
\end{align}
\end{subequations}
We have
$\max_{\mathbf{h}_{k} \in \mathbb{H}_{k}} \xi_{\mathrm{c},k}\big( \mathbf{h}_{k}, \widehat{g}_{\mathrm{c},k}, \widehat{u}_{\mathrm{c},k} \big) \geq \widehat{\xi}_{\mathrm{c},k}\big( \widehat{g}_{\mathrm{c},k}, \widehat{u}_{\mathrm{c},k} \big)$ and
$\max_{\mathbf{h}_{k} \in \mathbb{H}_{k}} \xi_{k}\big( \mathbf{h}_{k}, \widehat{g}_{k}, \widehat{u}_{k} \big) \geq \widehat{\xi}_{k}\big( \widehat{g}_{\mathrm{c},k}, \widehat{u}_{k} \big)$, as the maximum is lower-bounded by the average for any distribution of $\mathbf{h}_{k}$ (or equivalently $\widetilde{\mathbf{h}}_{k}$  given $\widehat{\mathbf{h}}_{k}$) defined over $\mathbb{H}_{k}$.
Combining this with \eqref{Eq_wc_Rates_WMSEs_LB}, it follows that
\begin{equation}
\label{Eq_R_hat_UB_2}
\widehat{R}_{\mathrm{c},k}  \leq 1 - \! \! \!  \min_{\widehat{u}_{\mathrm{c},k}, \widehat{g}_{\mathrm{c},k}} \! \!
\widehat{\xi}_{\mathrm{c},k}\big( \widehat{g}_{\mathrm{c},k}, \widehat{u}_{\mathrm{c},k} \big)
\ \text{and} \
\widehat{R}_{k} \leq 1 - \!   \min_{\widehat{u}_{k}, \widehat{g}_{k}}
\widehat{\xi}_{k}\big( \widehat{g}_{k}, \widehat{u}_{k} \big)
\end{equation}
where the minimizations assume the closed-from solutions
$\big(\widehat{g}_{\mathrm{c},k}^{\mathrm{MMSE}},\widehat{u}_{\mathrm{c},k}^{\mathrm{MMSE}}  \big)$ and
$\big(\widehat{g}_{k}^{\mathrm{MMSE}},\widehat{u}_{k}^{\mathrm{MMSE}}  \big)$, with
$\widehat{u}_{\mathrm{c},k}^{\mathrm{MMSE}} \triangleq \big( \widehat{\varepsilon}_{\mathrm{c},k}^{\mathrm{MMSE}} \big)^{-1}$ and
$\widehat{u}_{k}^{\mathrm{MMSE}} \triangleq \big( \widehat{\varepsilon}_{k}^{\mathrm{MMSE}} \big)^{-1}$.
By substituting this back into \eqref{Eq_R_hat_UB_2}, the upper-bounds in \eqref{Eq_R_hat_UB} are obtained.
\hfill $\blacksquare$
\section{Proof of Proposition \ref{Proposition_Cutting_Set_Conv}}
\label{Appendix_Proof_Proposition_Cutting_Set_Conv}
Consider the semi-infinite optimization problem
\begin{equation}
\begin{aligned}
\label{Eq_Opt_Semi_inf}
\min_{\mathbf{x}} & \ \ f_{0}(\mathbf{x}) \\
\text{s.t.} & \ \ f_{m}(\mathbf{x},t) \leq 0, \ \forall t \in \mathcal{T}_{m}, m \in \mathcal{M}
\end{aligned}
\end{equation}
where $\mathcal{M} \triangleq \{1,\ldots,M\}$, and $\mathcal{T}_{1}$,\ldots,$\mathcal{T}_{M}$ are compact infinite index sets
(or uncertainty regions) \cite{Lopez2007,Shapiro2009a}\footnote{$\mathbf{x}$ is the optimization variable here and should not be confused with the transmit signal in \eqref{Eq_yk} and \eqref{Eq_x}.}.
The cutting-set algorithm solves \eqref{Eq_Opt_Semi_inf} by solving a sequence of sampled problems.
The $i$th sampled problem is given by
\begin{equation}
\begin{aligned}
\label{Eq_Opt_Sampled}
\min_{\mathbf{x}} & \ \ f_{0}(\mathbf{x}) \\
\text{s.t.} & \ \ f_{m}(\mathbf{x},t) \leq 0, \ \forall t \in \mathcal{T}_{m}^{(i)}, m \in \mathcal{M}
\end{aligned}
\end{equation}
where  $\mathcal{T}_{m}^{(i)} \subset \mathcal{T}_{m} $ is a finite subset.
Let $\mathcal{F}^{(i)}$ be the feasible set of the $i$th problem, and $\bar{\mathbf{x}}^{(i)} \in \mathcal{F}^{(i)}$  be a feasible solution (not necessarily optimum).
We assume that $\mathcal{F}^{(1)}$ is compact,
$f_{0}(\cdot)$ and $f_{1}(\cdot,t),\ldots,f_{M}(\cdot,t)$ are continuously differentiable in $\mathbf{x} \in \mathcal{F}^{(1)}$, and the pessimization step is exact.
Under such assumptions, it follows from \cite[Section 5.2]{Mutapcic2009} that the iterates generated by the cutting-set algorithm converge to a feasible point of problem \eqref{Eq_Opt_Semi_inf}.
In particular, we have
\begin{equation}
\label{Eq_limit_feasibility}
f_{m}(\bar{\mathbf{x}},t)  \leq 0, \forall t \in \mathcal{T}_{m}, m \in \mathcal{M}
\end{equation}
where $\bar{\mathbf{x}}$ is a limit point of the algorithm\footnote{While global optimality of the optimization step is assumed in \cite{Mutapcic2009}, it is not necessary for the convergence of the algorithm and the feasibility of its limit point. This is also shown in the proof of \cite[Theorem 2.1]{Wu2005}.}.
Next, we show that if $\bar{\mathbf{x}}^{(i)}$ is a KKT point of \eqref{Eq_Opt_Sampled} for all $i$, then $\bar{\mathbf{x}}$ is a KKT point of \eqref{Eq_Opt_Semi_inf}.
The Lagrangian of \eqref{Eq_Opt_Sampled} is given by
\begin{equation}
\label{Eq_L_Sampled}
L(\mathbf{x},\bm{\lambda}^{(i)})  =  f_{0}(\mathbf{x}) +
\sum_{m =1}^{M} \sum_{t \in \mathcal{T}_{m}^{(i)}} \lambda_{m,t}^{(i)} f_{m}(\mathbf{x},t)
\end{equation}
where $\bm{\lambda}^{(i)} \triangleq \big\{ \lambda_{m,t}^{(i)} \mid t \in \mathcal{T}_{m}^{(i)}, m \in \mathcal{M} \big\}$ is the associated set of non-negative multipliers.
We define the discrete measures $\mu_{1}^{(i)},\ldots,\mu_{M}^{(i)}$ on $\mathcal{T}_{1}$,\ldots,$\mathcal{T}_{M}$ respectively such that
\begin{equation}
\mu_{m}^{(i)}(t)=
\begin{cases}
        \lambda_{m,t}^{(i)}, & \forall t \in \mathcal{T}_{m}^{(i)}  \\
        0 , & \forall t \in \mathcal{T}_{m} \setminus \mathcal{T}_{m}^{(i)} .
\end{cases}
\end{equation}
It follows that the  Lagrangian in \eqref{Eq_L_Sampled} can be expressed as
\begin{equation}
\label{Eq_L_integral}
L(\mathbf{x},\bm{\mu}^{(i)})  =  f_{0}(\mathbf{x}) +
\sum_{m =1}^{M} \int_{t \in \mathcal{T}_{m}}  f_{m}(\mathbf{x},t)\mathrm{d}\mu_{m}^{(i)}(t)
\end{equation}
where $\bm{\mu}^{(i)} \triangleq \big\{ \mu_{m}^{(i)} \mid m \in \mathcal{M} \big\}$.
Let $(\bar{\mathbf{x}}^{(i)},\bar{\bm{\mu}}^{(i)})$ denote the KKT solution of problem \eqref{Eq_Opt_Sampled} obtained in the $i$ iteration and suppose that some regularity condition holds\footnote{In particular, it is assumed that the Mangasarian-Fromovitz Constraint Qualification (MFCQ) holds at stationary points of \eqref{Eq_Opt_Semi_inf} and \eqref{Eq_Opt_Sampled}.}. The corresponding KKT optimality conditions are given by
\begin{subequations}
\label{Eq_KKT_samp}
\begin{align}
\label{Eq_KKT_samp_1}
&
\nabla_{\mathbf{x}}L(\bar{\mathbf{x}}^{(i)},\bar{\bm{\mu}}^{(i)})  = \mathbf{0} \\
\label{Eq_KKT_samp_2}
&
f_{m}(\bar{\mathbf{x}}^{(i)},t)  \leq 0, \forall t \in \mathcal{T}_{m}^{(i)}, m \in \mathcal{M} \\
\label{Eq_KKT_samp_3}
&
\bar{\mu}_{m}^{(i)} \geq 0, \forall  m \in \mathcal{M} \\
&
\label{Eq_KKT_samp_4}
\int_{t \in \mathcal{T}_{m}} f_{m}(\bar{\mathbf{x}}^{(i)},t) \mathrm{d}\bar{\mu}^{(i)}_{m}(t)  = 0 , \forall  m \in \mathcal{M}
\end{align}
\end{subequations}
where $\bar{\mu}_{m}^{(i)} \geq 0$ means that the measure is non-negative.

The sequence $\big\{ \bar{\mathbf{x}}^{(i)} \big\}_{i = 1}^{\infty}$ lies in the compact set $\mathcal{F}^{(1)}$, as $\mathcal{F}^{(1)} \supseteq \mathcal{F}^{(i)}$ for all $i$.
Hence, there exists a subsequence $\big\{ \bar{\mathbf{x}}^{(i_{r})} \big\}_{r = 1}^{\infty}$ converging to $\bar{\mathbf{x}}$.
The regularity condition implies that at each $\bar{\mathbf{x}}^{(i)}$, the set of KKT multipliers that satisfy \eqref{Eq_KKT_samp} is bounded \cite{Gauvin1977}.
Therefore, it is assumed without loss of generality that the subsequence $\big\{  \bar{\bm{\mu}}^{(i_{r})} \big\}_{r = 1}^{\infty}$ converges weakly to the accumulation point $\bar{\bm{\mu}}$.
Combining these observations with the continuity of the objective and constraint functions and their gradients implies that the solution
$(\bar{\mathbf{x}},\bar{\bm{\mu}})$ satisfies
\begin{subequations}
\label{Eq_KKT_limit}
\begin{align}
\label{Eq_KKT_limit_1}
&
\nabla_{\mathbf{x}}L(\bar{\mathbf{x}},\bar{\bm{\mu}}) = \mathbf{0} \\
\label{Eq_KKT_limit_2}
&
\bar{\mu}_{m} \geq 0, \forall  m \in \mathcal{M}\\
\label{Eq_KKT_limit_3}
&
\int_{t \in \mathcal{T}_{m}} f_{m}(\bar{\mathbf{x}},t) \mathrm{d}\bar{\mu}_{m}(t)  = 0 , \forall  m \in \mathcal{M}
\end{align}
\end{subequations}
where \eqref{Eq_KKT_limit_1} and \eqref{Eq_KKT_limit_3} can be shown using the same steps in the proof of \cite[Theorem 2.1]{Wu2005},
while \eqref{Eq_KKT_limit_2} follows from  \eqref{Eq_KKT_samp_3}.
Combining \eqref{Eq_KKT_limit} with \eqref{Eq_limit_feasibility} implies that  $(\bar{\mathbf{x}},\bar{\bm{\mu}})$ satisfies the KKT conditions of problem \eqref{Eq_Opt_Semi_inf}, and $\bar{\mathbf{x}}$ is a KKT point\footnote{The semi-infinite problem in \eqref{Eq_Opt_Semi_inf} has finite active constraints at KKT points. Hence, the measures $\bar{\mu}_{1},\ldots,\bar{\mu}_{M}$ have finite supports \cite{Lopez2007,Shapiro2009a}.}.
Since the iterates lie in a compact set, the result holds for any sequence of iterates generated by the algorithm.
Now, we observe that problems \eqref{Eq_Problem_R_RS} and \eqref{Eq_Problem_R_RS_Samp} are instances of problems \eqref{Eq_Opt_Semi_inf} and \eqref{Eq_Opt_Sampled} respectively, with continuously differentiable objective and constraint functions \cite{Shi2011,Razaviyayn2013b}.
Also, $\mathbf{P}$ lies in the compact set given by $\big\{ \mathbf{P} \mid  \mathrm{tr}\big(\mathbf{P}\mathbf{P}^{H}\big) \leq P_{\mathrm{t}} \big\}$.
The same holds for the rate variables $\bar{R}_{\mathrm{t}}$ and $\bar{\mathbf{c}}$, which belong to compact rate regions.
Hence, the feasible sets for \eqref{Eq_Problem_R_RS} and \eqref{Eq_Problem_R_RS_Samp} are compact, which completes the proof.
\hfill $\blacksquare$
\section{Proof of Proposition \ref{Proposition_AO_Stationary}}
\label{Appendix_Proof_Proposition_AO_Stationary}
The AO procedure described in Section \ref{subsection_optimization} is an instance of the Successive Convex Approximation (SCA) method in \cite[Section 2.1]{Razaviyayn2014}.
In particular, updating $(\bar{R}_{\mathrm{t}},\bar{\mathbf{c}},\mathbf{P})$ in each iteration corresponds to solving a convex approximation of \eqref{Eq_Problem_R_RS_Samp}, where the WMSEs in \eqref{Eq_Problem_WMSE_R_RS_Samp} approximate the rates around $\mathbf{P}^{(n-1)}$, which is obtained from the previous iteration.
Moreover, it can be shown that the conditions in \cite[Assumption 1]{Razaviyayn2014} are satisfied and Slater's condition holds for the convex approximated problem (see Sections 3.1.2 and 3.1.4 in \cite{Razaviyayn2014}).
Hence, it follows from \cite[Theorem 1]{Razaviyayn2014} that any limit point of the AO procedure is a KKT point of problem \eqref{Eq_Problem_R_RS_Samp}.
Since the iterates lie in a compact set (shown in Appendix \ref{Appendix_Proof_Proposition_Cutting_Set_Conv}), the convergence to the set of KKT points follows (see \cite[Corollary 1]{Razaviyayn2013}).
\hfill $\blacksquare$
\ifCLASSOPTIONcaptionsoff
  \newpage
\fi
\bibliographystyle{IEEEtran}
\bibliography{IEEEabrv,References}
\end{document}